\documentclass[
 reprint,
 superscriptaddress,
 nofootinbib,
 amsmath,amssymb,
 aps,prl
]{revtex4-2}

\usepackage{graphicx}
\usepackage{color}
\usepackage{natbib}
\usepackage{epsfig}
\usepackage{float}
\usepackage{xcolor}
\usepackage{soul}
\usepackage{comment}
\usepackage{amsmath,amssymb,amsfonts}
\usepackage{orcidlink}
\usepackage{subfigure}
\usepackage{hyperref}
\usepackage[normalem]{ulem}
\usepackage{longtable}

\DeclareUnicodeCharacter{2212}{\textendash}

\newcommand{\apjl}{Astrophys. J. Lett.}%
\newcommand{\aap}{Astron. Astrophys.}%
\newcommand{\mnras}{Mon. Not. Roy. Astron. Soc.}%
\newcommand{\physrep}{Physics Reports}
\newcommand{\nphysa}{Nuclear Physics A}
\newcommand{\physscr}{Physica Scripta}

\begin{document}

%\preprint{}

\title{Stellar properties indicating the presence of hyperons in neutron stars}



\author{Andreas Bauswein\,\orcidlink{0000-0001-6798-3572}}

\affiliation{GSI Helmholtzzentrum f{\"u}r Schwerionenforschung,
  Planckstra{\ss}e 1, D-64291 Darmstadt, Germany}
\affiliation{Helmholtz Forschungsakademie Hessen f\"ur FAIR (HFHF),
  GSI Helmholtzzentrum f\"ur Schwerionenforschung, Planckstra{\ss}e~1,
  D-64291 Darmstadt, Germany}

\author{Aristeidis Nikolaidis\,\orcidlink{0009-0000-1354-2288}}
\affiliation{GSI Helmholtzzentrum f{\"u}r Schwerionenforschung,
  Planckstra{\ss}e 1, D-64291 Darmstadt, Germany}
\affiliation{Department of Physics and Astronomy, Ruprecht-Karls-Universität Heidelberg, 
Im Neuenheimer Feld 226, 69120 Heidelberg, Germany}

\author{Georgios Lioutas\,\orcidlink{0000-0002-1434-3712}}
\affiliation{Heidelberger Institut für Theoretische Studien (HITS), Schloss-Wolfsbrunnenweg 35, 69118 Heidelberg, Germany}

\author{Hristijan Kochankovski\,\orcidlink{0000-0002-3391-7320}}
\affiliation{Departament de F\'{\i}sica Qu\`antica i Astrof\'{\i}sica and Institut de Ci\`encies del Cosmos, Universitat de Barcelona, Mart\'i i Franqu\`es 1, 08028, Barcelona, Spain}

\author{Prasanta Char\,\orcidlink{0000-0001-6592-6590}}
\affiliation{Departamento de F\'isica Fundamental and IUFFyM, Universidad de Salamanca, Plaza de la Merced S/N, E-37008 Salamanca, Spain}
\affiliation{Space Sciences, Technologies and Astrophysics Research (STAR) Institute, Universit\'e de Li\`ege, B\^at. B5a, 4000 Li\`ege, Belgium}

\author{Chiranjib Mondal}

\affiliation{Institut d'Astronomie et d'Astrophysique, CP-226, Universit\'{e} Libre de Bruxelles, 1050 Brussels, Belgium}

\author{Micaela Oertel\, \orcidlink{0000-0002-1884-8654}}
\affiliation{Observatoire astronomique de Strasbourg, CNRS, Universit\'e de Strasbourg, 11 rue de l'Université, 67000 Strasbourg, France}
\affiliation{LUX, CNRS, Observatoire de Paris, Universit\'e PSL, Universit\'e Paris Cit\'e, 5 place Jules Janssen, 92195 Meudon, France}

\author{Laura Tolos\,\orcidlink{0000-0003-2304-7496}}
\affiliation{Institute of Space Sciences (ICE, CSIC), Campus UAB, Carrer de Can Magrans, 08193 Barcelona, Spain}
\affiliation{Institut d'Estudis Espacials de Catalunya (IEEC), 08860 Castelldefels (Barcelona), Spain}

\author{Nicolas Chamel\,\orcidlink{0000-0003-3222-0115}}
\affiliation{Institut d'Astronomie et d'Astrophysique, CP-226, Universit\'{e} Libre de Bruxelles, 1050 Brussels, Belgium}
\author{Stephane Goriely\,\orcidlink{0000-0002-9110-941X}}
\affiliation{Institut d'Astronomie et d'Astrophysique, CP-226, Universit\'{e} Libre de Bruxelles, 1050 Brussels, Belgium}

%\thanks{}


\date{\today}

\begin{abstract}
We describe distinctive stellar features indicating the presence of hyperons in neutron stars as compared to purely nucleonic systems. A strongly negative curvature of the mass-radius relation $R(M)$ is characteristic of hyperons, which can be determined from measurements of neutron stars with three different masses. Similarly, a reduced second derivative of the tidal deformability as function of mass $\lambda(M)$ points to hyperonic degrees of freedom in NS matter. The slopes of such curves $R(M)$ and $\lambda(M)$ can distinguish a hyperonic equation of state from purely nucleonic models if they appear increased (decreased for $\lambda(M)$) relative to the maximum mass of neutron stars.
\end{abstract}

%\end{abstract}

%\keywords{}

\maketitle

% body of paper here - Use proper section commands References should be done
% using the \cite, \ref, and \label commands
\section{Introduction}
Non-nucleonic degrees of freedom such as hyperons are widely discussed as possible components of neutron star (NS) matter.  
Since the constituents of high-density matter and their interactions are not precisely known, the equation of state (EoS) of NSs is subject to uncertainties~\cite{glendenning1997compact,Haensel:2007yy,Lattimer2016,Baym2018,Raduta2022,Schaffner-Bielich:2020psc,Kumar2024}. The EoS of cold NSs in beta equilibrium is a unique relation between pressure and energy density that unambiguously determines the stellar structure like the mass-radius relation or the tidal deformability of stars in binaries through the relativistic stellar structure equations~\cite{Tolman1939,Oppenheimer1939,Hinderer2008,Binnington2009,Hinderer2010}. By measuring stellar parameters of NSs it is thus possible to probe the EoS and learn about the fundamental building blocks of high-density matter, e.g. the presence of hyperons. These efforts are linked to numerous theoretical and experimental studies to understand the interactions between hyperons and nucleons~\cite{Gal2016,nsac2023,nupecc2025}.

Many theoretical models of the EoS with different degrees of sophistication have been devised for purely nucleonic matter as well as for matter with hyperons~\cite{Compose2022}. The occurrence of hyperons is known to be associated with a softening of the EoS, which would generally result in a reduction of NS radii and the maximum mass of NSs~\cite{Burgio:2021vgk,Sedrakian:2022ata,Sedrakian2022,Tolos:2020aln,Chatterjee:2015pua,Vidana2018,Schaffner-Bielich:2020psc,Logoteta:2021iuy}. 
However, to the best of our knowledge there is to date no clear and distinctive \emph{quantitative} measure known based on which a discrimination between purely nucleonic and hyperon-admixed matter in NSs is possible employing measurements of stellar structure properties. 
Only a statistical approach considering the slope of the mass-radius relation has been discussed in Ref.~\cite{Ferreira2025}, where radii decreasing with mass in the range $1.2~M_\odot$ to $1.4~M_\odot$ point to a lower likelihood of hyperons being present in NSs. 
Statistical arguments have also been employed in~\cite{Huang2025}, and cooling observations may also provide valuable information on the composition of NSs~\cite{Page1990,Prakash1992,Page2006,Negreiros2018,Fischer2018,Grigorian2018,Raduta2018,Raduta2019,Potekhin2020,Fortin:2021umb,Anzuini2022,Sedrakian:2022ata}.

The capabilities of NS observations to determine stellar properties are continuously growing for instance with the X-ray timing measurements of NICER~\cite{Salmi2024,Vinciguerra2024,Dittmann2024,Rutherford2024} or with gravitational-wave observatories~\cite{Abbott2017,Abbott2018,Abbott2019,Abbott2020}, and several new instruments are projected to advance these efforts in the next decade~\cite{CE2017,Strobe2024,ETbluebook2025,eXTP2025}. But the current lack of a distinct feature of hyperons imprinted on the stellar structure parameters implies that even precise measurements of NS properties would not allow to infer the hyperon content of NSs and solve the ``hyperon puzzle'' (cf.~\cite{Lonardoni2015,Burgio:2021vgk,Sedrakian:2022ata,Sedrakian2022,Tolos:2020aln,Chatterjee:2015pua,Vidana2018,Schaffner-Bielich:2020psc,Logoteta:2021iuy,nsac2023,Vidana2025,nupecc2025}).

In this paper we propose to consider the curvature and the second derivative with respect to mass of the mass-radius curve and the mass-tidal deformability relation to identify the presence of hyperons in NSs or, generally, non-nucleonic degrees of freedom. Those parameters quantify the bending towards smaller NS radii with increasing mass and are found to be characteristically lowered if hyperons are present. We also find that the slope of the aforementioned relations at a fixed mass compared to the maximum mass of NSs can indicate hyperonic degrees of freedom. Observationally, these characteristics are accessible by considering several measurements of NSs with different masses. The stellar structure does not give a handle on which mechanisms and degrees of freedom are responsible for softening the EoS. Thus, ultimately, the properties discussed here can only indicate that non-nucleonic degrees of freedom soften the EoS so strongly that it is incompatible with purely nucleonic models. Non-nucleonic degrees of freedom appearing via a first-order phase transition (such as quarks in many models) result in a characteristic kink in the mass-radius relation, which has been discussed as potential signature for many years~\cite{Zdunik1987,glendenning1997compact,Haensel:2007yy,Schaffner-Bielich:2020psc}, and thus identifying a quantitative signature for a smoother occurrence of non-nucleonic degrees of freedom remains a major challenge.

\section{EoS sample}
We base our study on a very large set of microphysical EoS models for matter in beta equilibrium and at zero temperature. This sample includes 47 hyperonic models and 248 purely nucleonic EoSs (see appendix). These tables stem from three sources, to which we refer for more details. One subset consists of various models which we previously used in NS merger simulations~\cite{Blacker:2023opp,Kochankovski:2025lqc} and which were obtained through private communication or various webpages~\cite{Glendenning_1985, Muther_1987, Wiringa_1988, Goriely:2010bm, Lattimer_1991, Sugahara_1994, Toki_1995, Engvik_1996, Chabanat_1998, Akmal_1998, Alford_2005, Typel_2005, Lackey_2006, Read_2009, Hempel_2010, Typel_2010, Hempel_2012, Shen_2011, Steiner_2013, Banik_2014, Alvarez-Castillo_2016, Dexheimer_2017, Tolos_2017, Togashi_2017, Furusawa_2017, Schneider_2017, Marques_2017, Fortin_2018, Bombaci_2018, Schneider_2019, Logoteta_2021, Stone_2021, Sedrakian_2022, Du_2022, Kochankovski_2022, Kochankovski_2024, Tsiopelas_2024,Tong2025}. Another larger set of models is taken from the Compose repository~\cite{Compose2022}, where we include all baryonic NS matter models with a maximum mass of at least 1.9~$M_\odot$~\cite{Baym_1971, Deshmukh_1976, Friedrich_1986, Lee_1988, Rhoades_Brown_1989, Glendenning_1991, Reinhard_1995, Lalazissis_1997, Typel_1999, Bombaci_2000, Douchin_2001, Raduta_2002, Agrawal_2003, Gaitanos_2004, Long_2004, Lalazissis_2005, Maruyama_2005, van_Dalen_2005, Agrawal_2005, Danielewicz_2009, Dexheimer_2008, Dexheimer_2010, Sch_rhoff_2010, Gulminelli_2012, Goriely_2013, Xu_2013, Grill_2014, oertel2014hyperonsneutronstarmatter, Dexheimer_2015, Gulminelli_2015, Blumenfeld_2017, Wang_2017, Pais_2017, Gonzalez_Boquera_2018, Pearson_2018, Typel_2018, Dexheimer_2019, Perot_2019, Pearson_2020, Raduta_2020, Mondal_2020, Wei_2020, Taninah_2020, Allard_2021, Dexheimer_2021, Vi_as_2021, Hempel_2012, Hornick_2021, Raduta_2022, Xia_2022, Grams_2022, Clevinger_2022, Pearson_2022, Pradhan_2023, Scurto_2023, Char_2023, Alford_2023, Alford:2022bpp, Carreau_2019, Providencia_2019, Chen:2014sca, Pais:2016xiu, Shen:2020sec, Scurto:2024ekq}. % (as of February 2025). 
We supplement our sample with a selection of models for purely nucleonic~\cite{Char2023,Char2025} and hyperonic~\cite{Char2025b,Davis:2024nda} matter constructed from covariant density functional theory. 
Overall, our sample represents a large variety of models which are based on different theoretical frameworks (e.g. non-relativistic and relativistic density functional methods) 
and adjusted to different experimental data (properties of nuclear matter and nuclei). We include parametric models only if they are fitted to reproduce actual microphysical models and we refrain from using agnostic approaches to the EoS since they do not inform about the composition and thus the potential presence of hyperons. 
All EoS models reach a maximum NS mass of at least 1.9~$M_\odot$, which is somewhat below the limit imposed by heavy pulsars~\cite{Demorest2010ShapiroStar,Antoniadis:2013pzd,Romani:2022jhd,Cromartie2020RelativisticPulsar,Fonseca2016}, but it is advantageous to slightly enlarge the range of possible models (only 9 hyperonic and 8 purely nucleonic models have $M_\mathrm{max}<2.0~M_\odot$). Otherwise we do not employ any additional selection criteria for the EoS sample albeit various constraints might have already been imposed during the development of certain EoS models.
\begin{figure}%[tb]
	\includegraphics[width=8.9cm]{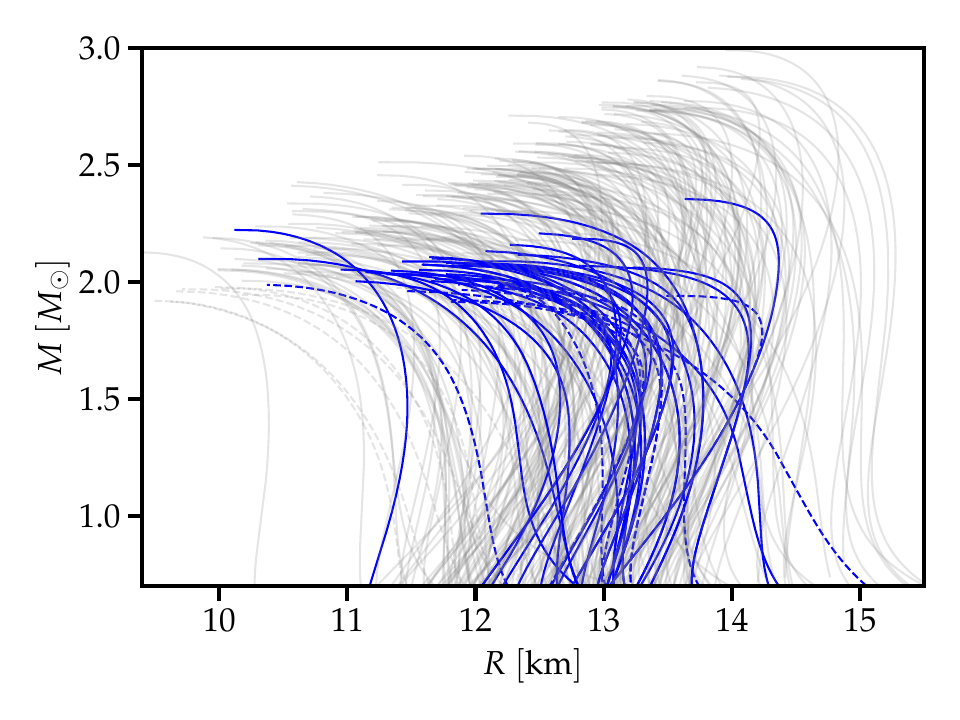}%Tovanalysis-discrete-newMR.py
	\caption{Mass-radius relations of NSs considered in this study for hyperonic EoSs (blue) and purely nucleonic EoSs (gray). Dashed curves indicate EoSs with $M_\mathrm{max}<2.0~M_\odot$.}\label{fig:mr}
\end{figure}

Since our sample includes many of the currently available microphysical EoS models, we assume that  our set is representative and covers sufficiently well the span of possible microphysical EoSs of purely nucleonic matter. 
The number of EoSs with hyperons is smaller because of the limited availability of such models, although the sample still covers a broad range with regards to the resulting stellar properties and, for instance, the EoSs from~\cite{Char2025b} have been chosen to represent the possible variations within the respective model. For our line of reasoning it suffices to show that hyperonic models exist which behave different from all possible purely nucleonic EoSs.

section{Radius curves}
For all EoSs we compute the circumferential radii $R$ and tidal deformability $\lambda$ as function of gravitational mass $M$ using the extended set of relativistic stellar structure equations~\cite{Tolman1939,Oppenheimer1939,Hinderer2008,Binnington2009,Hinderer2010}. $\lambda$ is given by $\frac{2}{3}k_2R^5$ with the tidal Love number $k_2$. 
All curves are provided in the appendix. In mass-radius relations (Fig.~\ref{fig:mr}) 
hyperonic models often exhibit the tendency to more strongly bend over to smaller radii with increasing mass. 
We intend to assess and quantify this effect as a characteristic indicator of hyperons in NSs. 

To this end we propose to consider the curvature of the curves $R(M)$. 
We compute $\kappa_R= {\frac{d^2 R}{d M^2}}{\left(1+\left(\frac{d R}{d M}\right)^2\right)^{-3/2}}$,
where we obtain the first and second derivative by finite differencing of the $R(M)$ curves\footnote{Using $R(M)$ instead of $M(R)$ avoids infinite derivatives on the stable branch of the mass-radius relation except for the configuration at $M_\mathrm{max}$. We employ a second order accurate finite differencing with variable step size.}.  
Geometrically $\kappa$ quantifies the inverse of the curvature radius of a curve. Using $G=c=1$ we express radii (and $\lambda$) in units of mass such that derivatives and $\kappa$ are dimensionless.

\begin{figure}%[tb]
	\includegraphics[width=8.9cm]{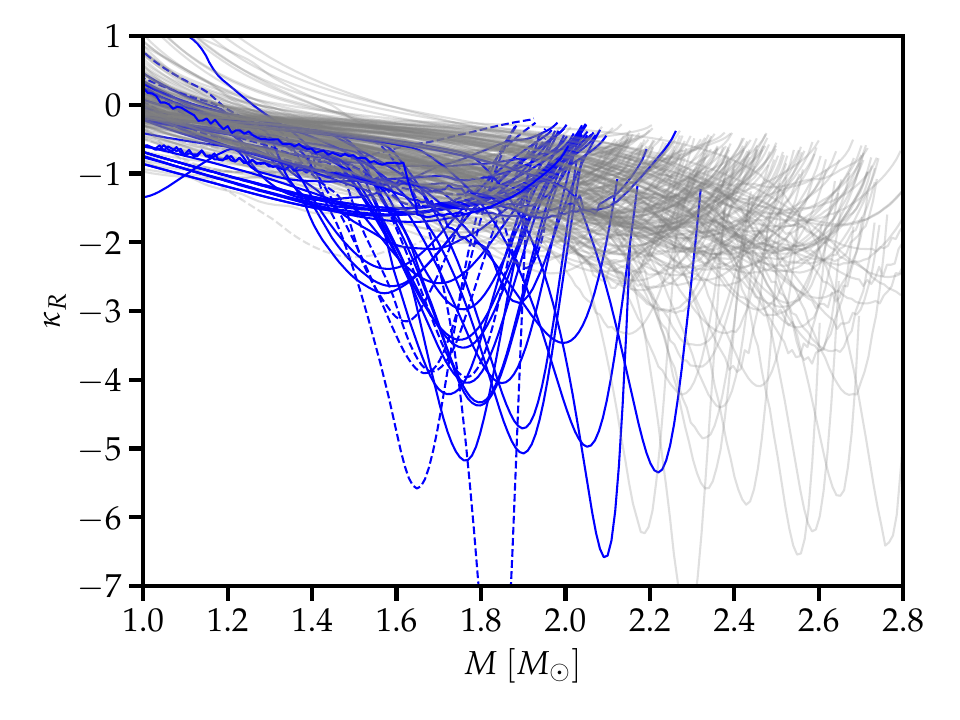}%Tovanalysis-discrete-kappa-M.py
	\caption{\label{fig:kappar} Curvature $\kappa_R$ as function of mass for purely nucleonic (gray) and hyperonic (blue) EoSs. Dashed curves indicate EoSs with $M_\mathrm{max}<2.0~M_\odot$. For numerical reasons the curves terminate slightly before $M_\mathrm{max}$.}
\end{figure}

Figure~\ref{fig:kappar} shows $\kappa_R$ as function of mass for hyperonic and purely nucleonic EoS models. As may already be obvious from the mass-radius relations, the curvature typically decreases with mass turning from positive to negative $\kappa_R$ and reaches a minimum to finally rise as $M$ goes to $M_\mathrm{max}$. 
The mass-radius relations of hyperonic models can feature particularly small, i.e.~strongly negative, curvatures in comparison to nucleonic EoSs especially for $1~M_\odot\lesssim M \lesssim 2~M_\odot$. Only some nucleonic models with $M_\mathrm{max}$ well above $2.0~M_\odot$ can yield similarly small $\kappa_R$ at larger masses above $\sim 2.0~M_\odot$. 

The small curvature for hyperonic EoSs is a consequence of the softening of the EoS at densities where hyperons occur. The mass where $\kappa_R$ starts to significantly drop approximately corresponds to the onset density of hyperon production. A strongly negative curvature is found for most hyperonic EoSs of our sample although there are a few exceptions (7 out of  47 hyperonic models do not reach below $\kappa_R=-1.5$). %Tovanalysis-discrete-kappa-M-play.py
Considering the ten hyperonic models with the smallest minima of the curvature, two models have $M_\mathrm{max}<2.0\,M_\odot$ while eight EoSs are compatible with the constraints from pulsar measurements.

\begin{figure}%[tb]
	\includegraphics[width=8.9cm]{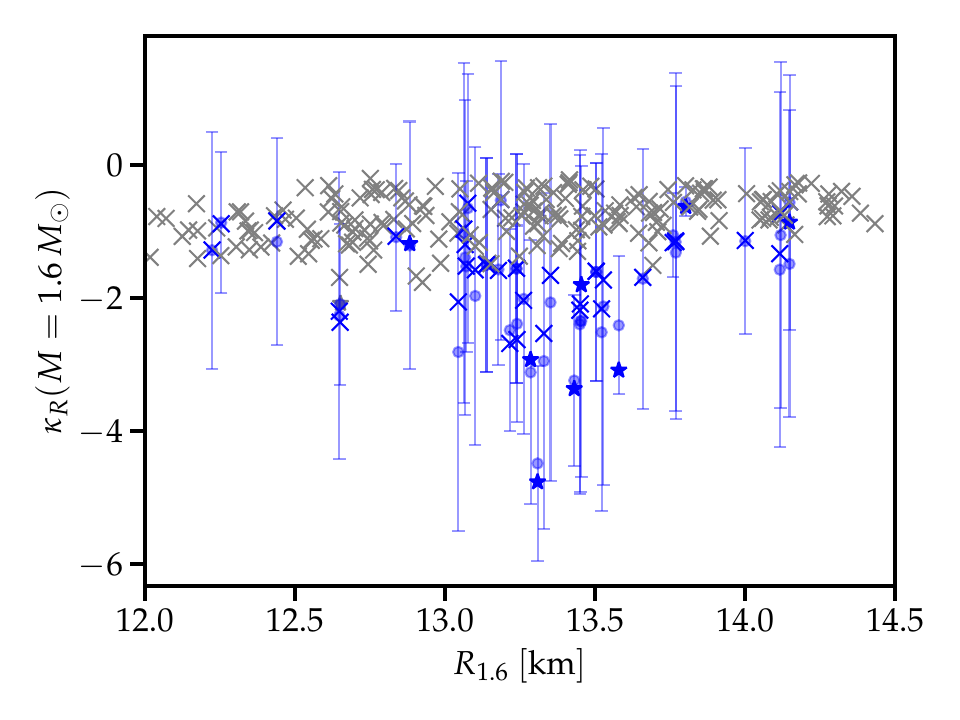}%{kappa-mref-error.pdf}%Tovanalysis-discrete-kappamref-errors.py
	\caption{Curvature $\kappa_R$ evaluated at a fixed mass of 1.6~$M_\odot$ as function the radius $R_{1.6}=R(1.6M_\odot)$ for purely nucleonic (gray) and hyperonic (blue) EoSs (crosses; asterisks for EoSs with $M_\mathrm{max}<2.0~M_\odot$). Radius measurements at $M=1.6M_\odot$ and $M=(1.6\pm0.25)M_\odot$ yield an estimate of $\kappa_R(1.6M_\odot)$ (dots). Error bars indicating corresponding precision for an assumed radius uncertainty of $\delta R=100$~meters (only for hyperonic models).}\label{fig:kappaerr} 
\end{figure}% 

This finding suggests to measure $\kappa_R$ in the range $1~M_\odot\lesssim M \lesssim 2~M_\odot$ to discriminate a hyperonic EoS from purely nucleonic EoSs. In particular in the range $1.4~M_\odot\lesssim M \lesssim 2.0~M_\odot$ there exist no nucleonic model reaching below $\kappa_R=-2.5$. The smallest $\kappa_R$ reached by nucleonic models depends on mass, and in fact the bulk of the nucleonic models does not reach below $\kappa_R=-1.5$ for $M<2.0~M_\odot$ (below $\kappa_R=-1.0$ for $M<1.6~M_\odot$). Note that $\kappa_R$ not dropping below -1.5 does however not exclude the presence of hyperons.

A measurement of $\kappa_R$ could be achieved in future by accurate radius measurements at three different masses allowing for a finite differencing estimate of the first and second derivatives of $R(M)$. We estimate the requirements for the identification of a hyperonic EoS through a determination of $\kappa_R$, where we illustrate the discussion by considering $\kappa_R$ at a fixed reference mass $M_\mathrm{ref}=1.6~M_\odot$. Analogue results can be obtained for $1.5~M_\odot\lesssim M_\mathrm{ref}\lesssim 2.1~M_\odot$. The crosses in Fig.~\ref{fig:kappaerr} display $\kappa_R(M=1.6\,M_\odot)$ as function of the radius $R_{1.6}$ of a $1.6~M_\odot$ NS demonstrating again the finding from Fig.~\ref{fig:kappar} that hyperonic models are distinguished by particularly small curvatures.

There are two sources of error to be considered when the curvature is extracted from measurements. (1) Computing the derivatives through finite differencing requires the measurement at three different masses with the ``discretization error'' increasing with the step size $\Delta M$, i.e. the difference between the masses where radii are measured. For simplicity we here assume a constant $\Delta M$ around $M_\mathrm{ref}=1.6~M_\odot$, i.e. measurements at 1.6~$M_\odot-\Delta M$, $1.6~M_\odot$ and $1.6~M_\odot+\Delta M$. 
(2) The radius measurements will only have a finite precision, which implies a corresponding uncertainty when the finite difference formulae are evaluated. For simplicity we adopt the same uncertainties $\delta R$ in all three measurements and neglect errors in the mass determinations, which we assume to be absorbed in the radius uncertainties.

\begin{figure}%[tb]
	\includegraphics[width=8.9cm]{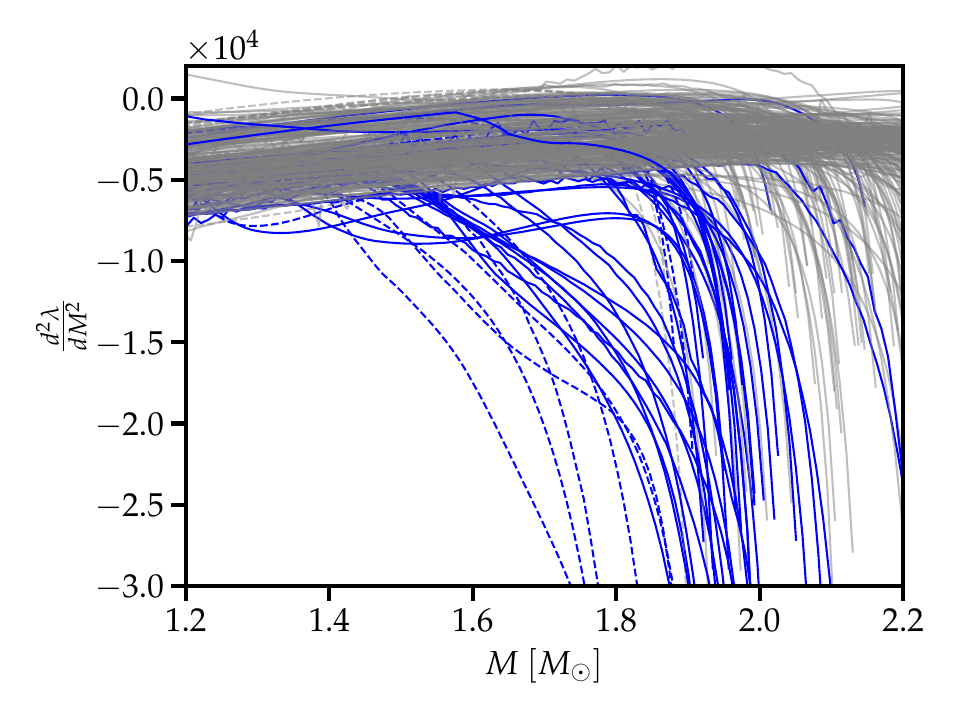}%Tovanalysis-discrete-secderivLambda-M.py
	\caption{Second derivative of $\lambda(M)$ with respect to mass $M$ for hyperonic EoSs (blue) and purely nucleonic EoSs (gray). Dashed curves indicate models with $M_\mathrm{max}<2.0~M_\odot$.}\label{fig:secderlam}
\end{figure}

The dots in Fig.~\ref{fig:kappaerr} show which $\kappa_R$ would be inferred for $\Delta M=0.25~M_\odot$ with second-order centered finite differencing. The difference between the crosses and dots quantifies the error associated to (1), and this ``discretezation error'' remains small for $\Delta M\leq 0.3~M_\odot$. For chosen $\Delta M$ and $\delta R$ we then draw error bars in Fig.~\ref{fig:kappaerr} illustrating the error stemming from (2). Those are obtained by a standard error propagation through the finite difference formulae. 
In this specific example in Fig.~\ref{fig:kappaerr} with $\Delta M=0.25~M_\odot$ around $M_\mathrm{ref}=1.6~M_\odot$, an accuracy of the radius measurements of about 100~meters is required to distinguish a hyperonic EoS meaning that the corresponding error bars of certain hyperonic models hardly overlap with purely nucleonic models. The error bar grows linearly with $\delta R$ but decreases with larger $\Delta M$. 

The accuracy of about 0.1~km is not yet available in current measurements, although on longer terms an order of magnitude improvement of the error might not be completely unrealistic~\cite{Chatziioannou2022,Pacilio2022,Iacovelli2023,Bandopadhyay2024,Finstad2023,Walker2024,Huez2025}. We also note that statistical approaches and the inclusion of more than three measurements may improve the determination of $\kappa_R$. 
The exact requirements for an identification of hyperons through $\kappa_R$ also depend on the considered reference mass. For instance, at masses above 1.6~$M_\odot$ the differences in $\kappa_R$ between hyperonic and nucleonic models become more pronounced (Fig.~\ref{fig:kappar}) but obtaining measurements with sufficiently large $\Delta M$ might be less likely as one approaches $M_\mathrm{max}$.

$\frac{d^2R}{dM^2}$ at a given mass $1.5~M_\odot\lesssim M_\mathrm{ref}\lesssim2.0~M_\odot$ versus $R(M_\mathrm{ref})$ may also be an interesting indicator for hyperons, where several hyperonic models tend to lie at lower values than purely nucleonic EoSs (similar to Fig.~\ref{fig:kappaerr} but less pronounced, see appendix).

\section{Tidal deformability curves}
We now discuss the tidal deformability as indicator of hyperonic EoSs. 
We remark that the occurrence of features of hyperonic models in the first and second derivative with respect to mass and the curvature $\kappa$  depends on whether one considers $\lambda(M)$ or $\Lambda(M)=\lambda(M)/M^5$ because the specific effects of hyperons can cancel each other in different terms.

\begin{figure}%[tb]
	\includegraphics[width=8.9cm]{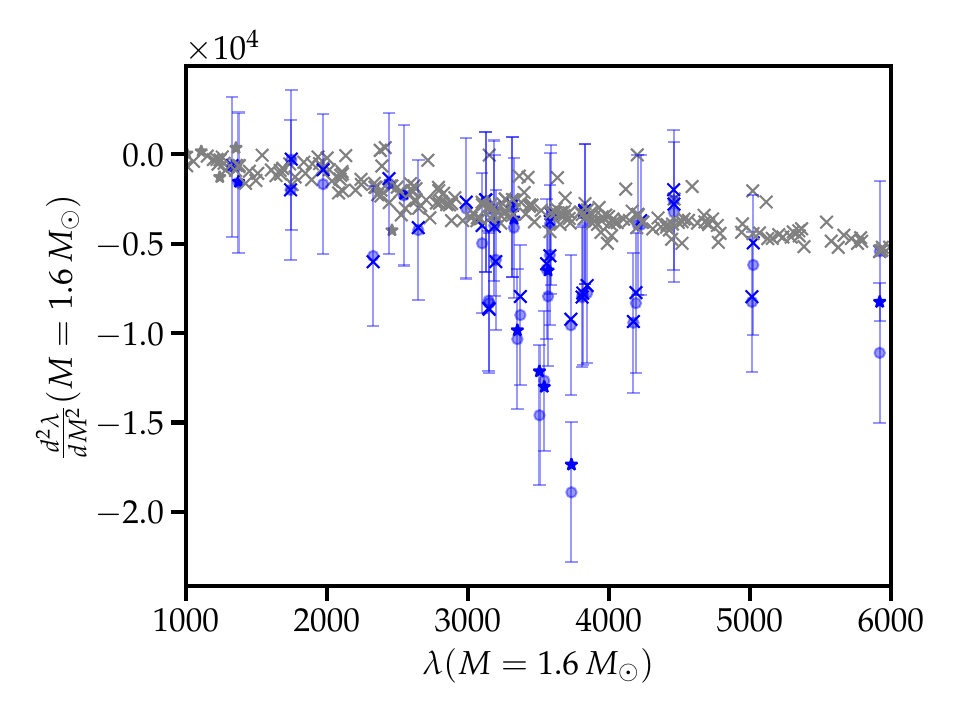}%{secderivLambda-M-mref-errors.pdf}%Tovanalysis-discrete-secderivLambda-M-mref-errors.py
	\caption{Second derivative of $\lambda(M)$ with respect to mass $M$ at fixed mass $M=1.6~M_\odot$ as function of $\lambda(M=1.6~M_\odot)$. Symbols and colors have same meaning as in Fig.~\ref{fig:kappaerr}. See main text for details.}\label{fig:secderlammreferror}
\end{figure}

The imprint of hyperons becomes distinct and clear if one considers the second derivative. This occurs very prominently in Fig.~\ref{fig:secderlam} showing the second derivative $\frac{d^2\lambda}{dM^2}(M)$, where hyperonic models yield particularly small values for $1.4~M_\odot\lesssim M \lesssim 1.85~M_\odot$. At larger masses, nucleonic models can feature similarly small $\frac{d^2\lambda}{dM^2}$ as all models exhibit a decrease of $\frac{d^2\lambda}{dM^2}$ towards $M_\mathrm{max}$.

The potential of this signature is further demonstrated in Fig.~\ref{fig:secderlammreferror} (analogue to Fig.~\ref{fig:kappaerr}). We provide an uncertainty estimate for $\frac{d^2\lambda}{dM^2}$ at $M_\mathrm{ref}=1.6~M_\odot$ with $\Delta M=0.25~M_\odot$. 
For the figure we assume $\delta \lambda=100$, i.e. a determination of $\lambda$ within a few per cent, which seems sufficient to discriminate hyperonic EoSs from purely nucleonic EoSs. Depending on the exact EoS, even larger $\delta \lambda$ could still yield evidence for hyperons in NSs noting that the uncertainty grows linearly with $\delta \lambda$. The difference between the crosses and the dots again visualizes the ``discretization error'', which is still acceptable for $\Delta M=0.25~M_\odot$. Similar figures can be obtained for other choices of the reference mass $1.4~M_\odot\lesssim M_\mathrm{ref}\lesssim 1.85~M_\odot$. For $\kappa_\lambda(M)$, i.e. the inverse curvature radius in $\lambda(M)$, we do not find features indicative of hyperonic EoSs.

It is similarly possible to consider $\Lambda(M)$, where hyperonic models do not stand out prominently in relations like $\frac{d^2\Lambda}{dM^2}(M)$ or $\kappa_\Lambda(M)$. Evaluating those quantities at fixed reference mass $1.6~M_\odot \lesssim M_\mathrm{ref} \lesssim 2.0~M_\odot$ versus $\Lambda(M_\mathrm{ref})$, some hyperonic models yield a characteristically reduced $\kappa_\Lambda(M_\mathrm{ref})$ compared to purely nucleonic EoSs, whereas $\frac{d^2\Lambda}{dM^2}(M_\mathrm{ref})$ is elevated (but some nucleonic models are similarly enhanced). For those hyperonic models, measurements of $\Lambda$ within a few per cent are sufficient to identify this signature of hyperons. However, in comparison to $\frac{d^2\lambda}{dM^2}$ the signatures are not as pronounced and especially for $\frac{d^2\Lambda}{dM^2}$ several purely nucleonic models overlap (see appendix). 

\begin{figure}%[tb]
    \includegraphics[ trim={0 0 0 0},width=8.0cm]{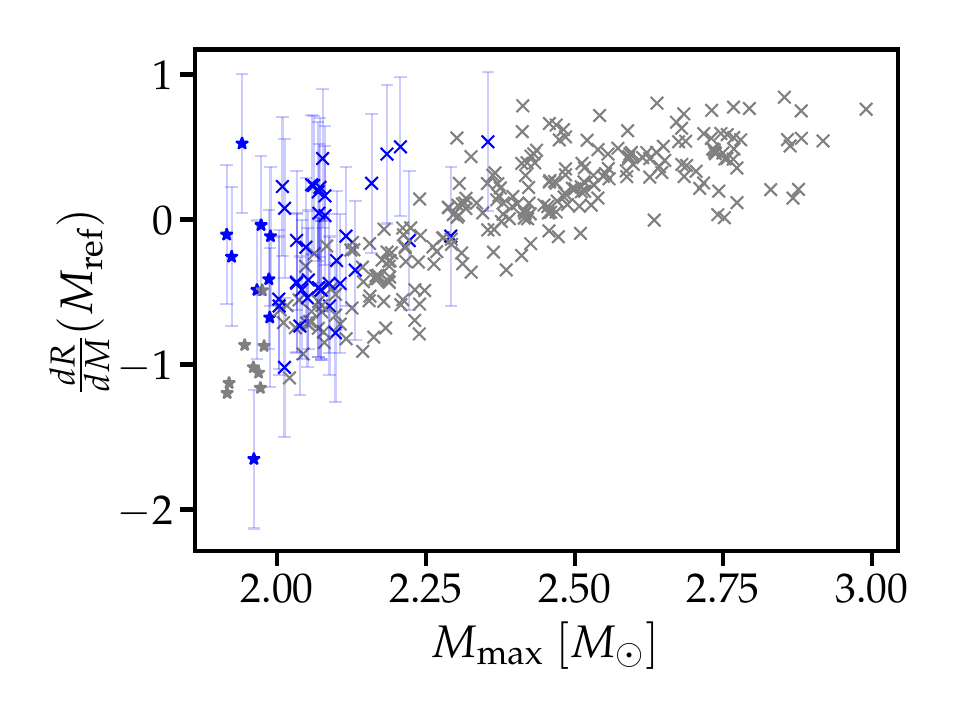}\\%Tovanalysis-discrete-slopeR-M-mref-Mmax-errors.py
    \includegraphics[ trim={0 0 0 0.8cm},width=8.0cm]{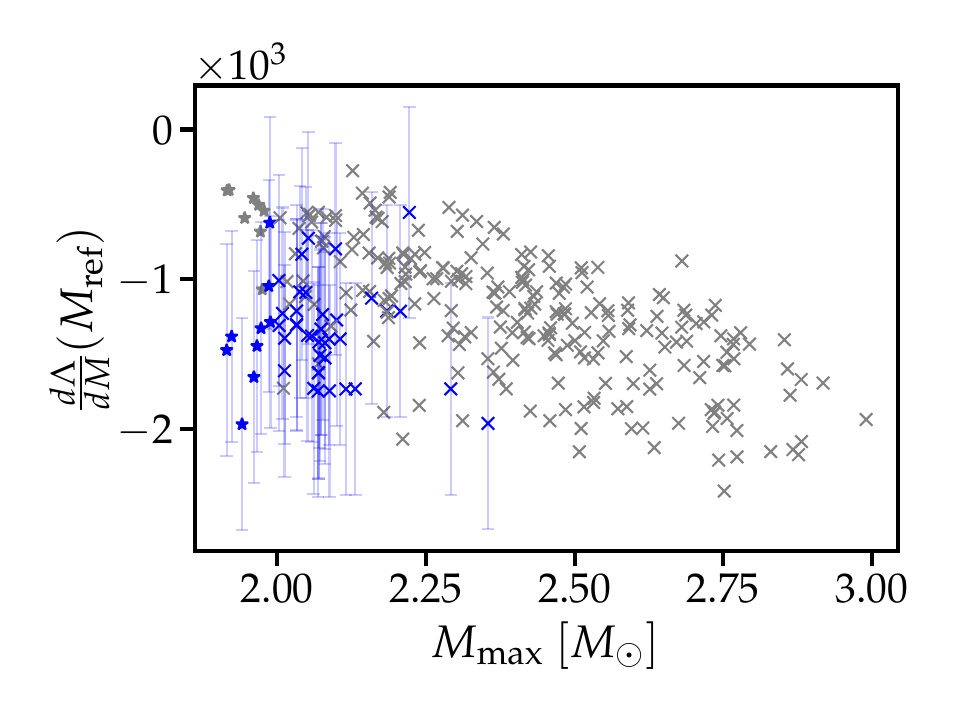}%Tovanalysis-discrete-slopecapLambda-M-mref-Mmax-errors.py
	\caption{Slope of $R(M)$ (top) and $\Lambda(M)$ (bottom) at a fixed mass of $M_\mathrm{ref}=1.6~M_\odot$ as function of the maximum mass of NS. Symbols and colors have the same meaning as in Fig.~\ref{fig:kappaerr} with error bars computed for the values mentioned in the main text}.\label{fig:slopes}
\end{figure}

\section{Slopes}
Finally, we highlight the slopes $\frac{dR}{dM}$ and $\frac{d\Lambda}{dM}$ as indicator for hyperons (cf.~\cite{Xie2020,Han2020,Zhao2020,Ferreira2024,Li2024,Cai2025,Ferreira2025,Tang2025,Kalita2025}. Reference~\cite{Ferreira2025} argued that $\frac{dR}{dM}<0$ in the range $1.2~M_\odot\leq M\leq 1.4~M_\odot$ would favor a nucleonic model employing statistical considerations without noting a distinctive feature for hyperonic EoSs. The slope of $R(M)$ or similarly of $\lambda(M)$ and $\Lambda(M)$ at a fixed mass on its own is not very conclusive with regards to the presence of hyperons because nucleonic models overlap over the whole range of hyperonic models (see appendix; with $\frac{d\lambda}{dM}$ appearing slightly more promising).

We however point out that comparing the slopes at a fixed mass to the maximum mass, some specific hyperonic models stand out clearly. Figure~\ref{fig:slopes} shows $\frac{dR}{dM}$ and $\frac{d\Lambda}{dM}$ evaluated at $M_\mathrm{ref}=1.6~M_\odot$ as function of $M_\mathrm{max}$ demonstrating that an increased (decreased) slope in $\frac{dR}{dM}$ ($\frac{d\Lambda}{dM}$) compared to $M_\mathrm{max}$ is a distinctive feature of hyperonic models. For $\frac{d\Lambda}{dM}$ there are however two nucleonic models with similarly small slope. For $\frac{d\lambda}{dM}$ hyperonic EoSs do not occur distinguished. Similar findings hold for $1.2~M_\odot\lesssim M_\mathrm{ref}\lesssim 1.8~M_\odot$. At first glance, an increased slope of $R(M)$ may seem counterintuitive as indicator of hyperons as one would associate it with a certain stiffness of the EoS. In those models hyperons soften the EoS at higher densities and thus reduce $M_\mathrm{max}$ more strongly than any purely nucleonic model with the same slope.

Measuring a slope can be assessed by assuming measurements for NSs with two masses separated by $\Delta M$. Employing second-order finite differencing, the uncertainty in the slope is given by $\sqrt{2}\frac{\delta R}{\Delta M}$ and $\sqrt{2}\frac{\delta \Lambda}{\Delta M}$, respectively, neglecting the discretization error, which is very small for $\Delta M\sim0.2~M_\odot$ and $M_\mathrm{ref}\sim1.6~M_\odot$. The scales in Fig.~\ref{fig:slopes} suggest that $\delta R\sim 0.1$~km or $\delta \Lambda\sim 100$ (about ten per cent level) is sufficient with $\Delta M=0.2~M_\odot$ to identify hyperons. While these demands on measurements look relatively promising especially for the tidal deformability, the identification of hyperons via these features also requires a solid upper bound on $M_\mathrm{max}$, which may not be easily available.

\section{Discussion and outlook}
We describe stellar features which allow to quantitatively distinguish a hyperonic EoS from purely nucleonic EoSs. Requirements for corresponding measurements are generally challenging but not completely out of reach~\cite{Chatziioannou2022,Pacilio2022,Iacovelli2023,Bandopadhyay2024,Finstad2023,Walker2024,Huez2025}. In particular, the required precision for measurements of the tidal deformability of a few per cent might be feasible with the next generation of GW detectors~\cite{CE2017,ETbluebook2025} although this precision has to be achieved for high-mass NSs. Future work should thus explore the capabilities of new instruments~\cite{CE2017,Strobe2024,ETbluebook2025,eXTP2025} and the potential of statistical methods including for instance more than three measurements.

The effects described in this study are generally associated with a softening of the EoS which goes beyond what a purely nucleonic EoS may attain, and we did not explicitly consider other non-nucleonic degrees of freedom (apart from $\Delta$ resonances being considered in some of the hyperonic models). A transition to quark matter might mimic the behavior of hyperonic models and would thus need to be further discriminated by additional information, e.g. from experiments like heavy-ion collisions, NS cooling or NS mergers, e.g.~\cite{Gal2016,Durante2019,Kumar2024,Friman:2011zz,Fischer2018,Page1990,Prakash1992,Page2006,Negreiros2018,Grigorian2018,Raduta2018,Raduta2019,Potekhin2020,Fortin:2021umb,Anzuini2022,Sedrakian:2022ata,Most2019,Bauswein2019,Blacker:2023opp,Kochankovski:2025lqc}. If quark matter occurs via a first-order phase transition (as opposed to the rather smooth appearance of hyperons), a characteristic kink or even discontinuity in the mass-radius relation is present~\cite{glendenning1997compact,Haensel:2007yy,Schaffner-Bielich:2020psc}. Future work should also elaborate on which particular properties of hyperonic matter lead to pronounced features, while some hyperonic models do not exhibit prominent differences compared to purely nucleonic EoSs. A refined study may disregard some of the nucleonic models of our sample by imposing further experimental or theoretical constraints, by which hyperonic EoSs may stand out even more clearly in the discussed relations.

\begin{acknowledgements}
  We thank David Blaschke, Gabriel Martinez-Pinedo and Angels Ramos for helpful discussions, and Hui Tong for providing EoS tables. A.B. and A.N. acknowledge support by the European Union through ERC Synergy Grant HeavyMetal no. 101071865. A.B. acknowledges support by Deutsche Forschungsgemeinschaft (DFG, German Research Foundation) through Project-ID 279384907 -- SFB 1245 (subproject B07). G.L acknowledges support by the Klaus Tschira Foundation. This project has received funding from the European Union's Horizon 2020 research and innovation programme under the Marie Skłodowska-Curie grant agreement No. 101034371. PC acknowledges the support from the European Union's HORIZON MSCA-2022-PF-01-01 Programme under Grant Agreement No. 101109652, project ProMatEx-NS. C.M. acknowledges partial support from the Fonds de la Recherche Scientifique (FNRS, Belgium) and the Research Foundation Flanders (FWO, Belgium) under the EOS Project nr O022818F and O000422. M.O. acknowledges financial support of the Agence Nationale de la Recherche (ANR) under contract ANR-22-CE31-0001-01. L.T. acknowledges support from the program Unidad de Excelencia María de Maeztu CEX2020-001058-M, from the project PID2022-139427NB-I00 financed by the Spanish MCIN/AEI/10.13039/501100011033/FEDER, UE (FSE+),  from  the Grant CIPROM 2023/59 of Generalitat Valenciana and from CRC-TR 211 ’Strong-interaction matter under extreme conditions’, Project No. 315477589 - TRR 211 by the Deutsche Forschungsgemeinshaft. S.G. and N.C acknowledge financial support from F.R.S.-FNRS (Belgium). This work was supported by the Fonds de la Recherche Scientifique - FNRS and the Fonds Wetenschappelijk Onderzoek - Vlaanderen (FWO) under the EOS Project No O000422.  
\end{acknowledgements}

\appendix
\section{EoS models and additional plots}\label{app}
We list all EoS models used in this study in Tab.~\ref{tab:eos_table_properties} along with the respective references and some stellar parameters.

%\clearpage

% \begin{longtable}{p{0.2\linewidth} c c c c p{0.3\linewidth}}
\begin{center}
\begin{longtable}{p{0.35\linewidth}p{0.08\linewidth}p{0.08\linewidth}ccp{0.18\linewidth}} 
\caption{\parbox[t]{0.4\textwidth}{EoS models employed in this work. Columns provide EoS acronym, maximum mass, ($M_{\rm max}$), radius and dimensionless tidal deformability at 1.6 $M_\odot$, ($R_{1.6}$ and $\Lambda_{1.6}$, respectively), presence of hyperons and references to the EoS.\newline} \\[1.0em]}
\label{tab:eos_table_properties} \\

    EoS & $M_{\rm max}$ \rm{[$M_\odot$]} & $R_{1.6}$ $\rm{[km]}$ & $\Lambda_{1.6}$ & Hyperons & Reference \\
    \hline
    \endfirsthead
    \multicolumn{6}{c}%
    {{\bfseries Table \thetable\ (continued)}} \\
    EoS & $M_{\rm max}$ $\rm{[M_\odot]}$ & $R_{1.6}$ $\rm{[km]}$ & $\Lambda_{1.6}$ & Hyperons & Reference \\
    \hline
    \endhead
    DDLS(30)-N & 2.48 & 13.06 & 325.4 & No & \cite{Tsiopelas_2024} \\
    DDLS(30)-Y & 2.01 & 13.04 & 321.7 & Yes & \cite{Tsiopelas_2024} \\
    DDLS(50)-N & 2.47 & 13.32 & 327.8 & No & \cite{Tsiopelas_2024} \\
    DDLS(50)-Y & 1.99 & 13.29 & 319.7 & Yes & \cite{Tsiopelas_2024} \\
    DDLS(70)-N & 2.46 & 13.64 & 352.1 & No & \cite{Tsiopelas_2024} \\
    DDLS(70)-Y & 1.97 & 13.58 & 337.9 & Yes & \cite{Tsiopelas_2024} \\
    Bonn1 & 2.23 & 12.44 & 188.3 & No & \cite{Tong2025, Baym_1971} \\
    Bonn2 & 2.04 & 12.44 & 188.3 & Yes & \cite{Tong2025, Baym_1971} \\
    DNS & 2.09 & 14.0 & 403.2 & Yes & \cite{Dexheimer_2017} \\
    DSH\,Fiducial & 2.17 & 11.67 & 121.3 & No & \cite{Du_2022} \\
    DSH\,Large\,$\mathrm{M}_\mathrm{max}$ & 2.21 & 12.67 & 222.9 & No & \cite{Du_2022} \\
    DSH\,Large\,SL & 2.16 & 11.66 & 110.0 & No & \cite{Du_2022} \\
    DSH\,Large\,R & 2.12 & 12.34 & 175.3 & No & \cite{Du_2022} \\
    DSH\,Small\,SL & 2.18 & 11.69 & 138.4 & No & \cite{Du_2022} \\
    DSH\,Smaller\,R & 2.14 & 11.26 & 96.0 & No & \cite{Du_2022} \\
    FSU2H$^*$L & 1.92 & 13.31 & 356.2 & Yes & \cite{Kochankovski_2024} \\
    FSU2H$^*$U & 2.06 & 13.35 & 366.9 & Yes & \cite{Kochankovski_2024} \\
    FSU2H$^*$ & 2.01 & 13.33 & 355.9 & Yes & \cite{Kochankovski_2022} \\
    FSU2R & 2.06 & 12.93 & 275.2 & No & \cite{Tolos_2017} \\
    FTNS & 2.22 & 11.46 & 126.7 & Yes & \cite{Togashi_2017,Furusawa_2017} \\
    LPB & 2.1 & 12.26 & 166.7 & Yes & \cite{Logoteta_2021,Bombaci_2018} \\
    QMC-A & 1.99 & 12.88 & 242.8 & Yes & \cite{Stone_2021} \\
    SRO(SLy4) & 2.06 & 11.57 & 116.4 & No & \cite{Schneider_2017,Chabanat_1998} \\
    BHBLP & 2.1 & 13.26 & 305.1 & Yes & \cite{Banik_2014} \\
    DD2 & 2.42 & 13.32 & 317.5 & No & \cite{Hempel_2010,Typel_2010} \\
    DD2Y & 2.03 & 13.24 & 300.3 & Yes & \cite{Marques_2017,Fortin_2018} \\
    DD2F & 2.08 & 12.24 & 165.6 & No & \cite{Alvarez-Castillo_2016,Typel_2005,Typel_2010} \\
    WFF1 & 2.13 & 10.37 & 60.7 & No & \cite{Wiringa_1988} \\
    WFF2 & 2.19 & 11.07 & 92.8 & No & \cite{Wiringa_1988} \\
    BSK20 & 2.17 & 11.66 & 129.5 & No & \cite{Goriely:2010bm} \\
    BSK21 & 2.28 & 12.55 & 223.3 & No & \cite{Goriely:2010bm} \\
    LS220 & 2.04 & 12.54 & 203.6 & No & \cite{Lattimer_1991} \\
    LS375 & 2.71 & 13.84 & 446.0 & No & \cite{Lattimer_1991} \\
    GS1 & 2.75 & 14.96 & 649.2 & No & \cite{Shen_2011} \\
    GS2 & 2.09 & 13.44 & 308.5 & No & \cite{Shen_2011} \\
    ALF2 & 1.98 & 12.65 & 234.9 & No & \cite{Alford_2005,Read_2009} \\
    AP3 & 2.37 & 11.97 & 159.3 & No & \cite{Akmal_1998,Read_2009} \\
    ENG & 2.24 & 11.91 & 153.3 & No & \cite{Engvik_1996,Read_2009} \\
    GNH3 & 1.96 & 13.8 & 317.4 & Yes & \cite{Glendenning_1985,Read_2009} \\
    H4 & 2.01 & 13.76 & 341.7 & Yes & \cite{Lackey_2006,Read_2009} \\
    MPA1 & 2.46 & 12.46 & 213.9 & No & \cite{Muther_1987,Read_2009} \\
    SFHO & 2.06 & 11.8 & 131.4 & No & \cite{Steiner_2013} \\
    SFHOY & 1.99 & 11.8 & 131.1 & Yes & \cite{Marques_2017,Fortin_2018} \\
    SFHX & 2.13 & 12.02 & 167.3 & No & \cite{Steiner_2013} \\
    SkAPR & 2.03 & 12.37 & 176.3 & No & \cite{Lattimer_1991,Schneider_2017,Schneider_2019} \\
    TM1 & 2.21 & 14.43 & 488.1 & No & \cite{Sugahara_1994,Hempel_2012} \\
    TMA & 2.01 & 13.69 & 376.3 & No & \cite{Hempel_2012,Toki_1995} \\
    ABHT(QMC-RMF1) & 1.95 & 11.66 & 118.4 & No & \cite{Baym_1971,Grill_2014,Alford_2023,Alford:2022bpp} \\
    ABHT(QMC-RMF2) & 2.04 & 11.90 & 142.8 & No & \cite{Baym_1971,Grill_2014,Alford_2023,Alford:2022bpp} \\
    ABHT(QMC-RMF3) & 2.15 & 12.15 & 159.6 & No & \cite{Baym_1971,Grill_2014,Alford_2023,Alford:2022bpp} \\
    ABHT(QMC-RMF4) & 2.21 & 12.35 & 201.6 & No & \cite{Baym_1971,Grill_2014,Alford_2023,Alford:2022bpp} \\
    APR & 2.19 & 11.26 & 100.7 & No & \cite{Baym_1971,Akmal_1998,Douchin_2001} \\
    BBB(BHF-BBB2) & 1.92 & 10.86 & 72.4 & No & \cite{Baym_1971,Bombaci_2000,Douchin_2001} \\
    BL(chiral)\,w.\,un.\,crust & 2.08 & 12.10 & 147.1 & No & \cite{Bombaci_2018,Carreau_2019,Davis:2024nda} \\
    CMGO(GDFM-II) & 2.30 & 13.89 & 431.1 & No & \cite{Char_2023,Carreau_2019} \\
    CMGO(GDFM-I) & 2.31 & 12.80 & 233.8 & No & \cite{Char_2023,Carreau_2019} \\
    DNS(CMF) & 2.07 & 13.51 & 365.3 & Yes & \cite{Dexheimer_2008,Sch_rhoff_2010,Dexheimer_2015,Gulminelli_2015} \\
    DS(CMF-1-Hybrid) & 2.07 & 13.50 & 365.3 & Yes & \cite{Dexheimer_2008,Dexheimer_2010,Dexheimer_2017,Dexheimer_2019,Dexheimer_2021,Clevinger_2022} \\
    DS(CMF-2-Hybrid) & 2.13 & 13.66 & 401.1 & Yes & \cite{Dexheimer_2008,Dexheimer_2010,Dexheimer_2017,Dexheimer_2019,Dexheimer_2021,Clevinger_2022} \\
    DS(CMF-3-Hybrid) & 2.00 & 13.07 & 285.1 & Yes & \cite{Dexheimer_2008,Dexheimer_2010,Dexheimer_2017,Dexheimer_2019,Dexheimer_2021,Clevinger_2022} \\
    DS(CMF-4-Hybrid) & 2.05 & 13.18 & 303.9 & Yes & \cite{Dexheimer_2008,Dexheimer_2010,Dexheimer_2017,Dexheimer_2019,Dexheimer_2021,Clevinger_2022} \\
    DS(CMF-5-Hybrid) & 2.07 & 13.14 & 298.3 & Yes & \cite{Dexheimer_2008,Dexheimer_2010,Dexheimer_2017,Dexheimer_2019,Dexheimer_2021,Clevinger_2022} \\
    DS(CMF-6-Hybrid) & 2.11 & 13.24 & 316.3 & Yes & \cite{Dexheimer_2008,Dexheimer_2010,Dexheimer_2017,Dexheimer_2019,Dexheimer_2021,Clevinger_2022} \\
    DS(CMF-7-Hybrid) & 2.07 & 13.14 & 298.3 & Yes & \cite{Dexheimer_2008,Dexheimer_2010,Dexheimer_2017,Dexheimer_2019,Dexheimer_2021,Clevinger_2022} \\
    DS(CMF-8-Hybrid) & 2.09 & 13.24 & 316.3 & Yes & \cite{Dexheimer_2008,Dexheimer_2010,Dexheimer_2017,Dexheimer_2019,Dexheimer_2021,Clevinger_2022} \\
    GDTB(DDHdelta) & 2.16 & 12.65 & 253.0 & No & \cite{Gaitanos_2004, Grill_2014, Douchin_2001} \\
    GM(GM1) & 2.39 & 14.11 & 425.8 & No & \cite{Glendenning_1991,Douchin_2001} \\
    GMSR(BSK14) & 1.92 & 10.88 & 72.3 & No & \cite{Grams_2022} \\
    GMSR(DHSL59) & 2.43 & 12.47 & 199.7 & No & \cite{Grams_2022} \\
    GMSR(DHSL69) & 2.41 & 12.51 & 200.2 & No & \cite{Grams_2022} \\
    GMSR(F0) & 2.07 & 11.55 & 117.2 & No & \cite{Grams_2022} \\
    GMSR(H1) & 2.29 & 11.51 & 129.1 & No & \cite{Grams_2022} \\
    GMSR(H2) & 2.31 & 11.71 & 144.1 & No & \cite{Grams_2022} \\
    GMSR(H3) & 2.30 & 12.04 & 169.4 & No & \cite{Grams_2022} \\
    GMSR(H4) & 2.34 & 11.87 & 156.5 & No & \cite{Grams_2022} \\
    GMSR(H5) & 2.38 & 12.17 & 180.5 & No & \cite{Grams_2022} \\
    GMSR(H7) & 2.51 & 12.85 & 251.1 & No & \cite{Grams_2022} \\
    GMSR(LNS5) & 1.98 & 11.41 & 105.9 & No & \cite{Grams_2022} \\
    GMSR(SLy5) & 2.10 & 11.63 & 120.1 & No & \cite{Grams_2022} \\
    GPPVA(DD2) & 2.42 & 13.22 & 311.6 & No & \cite{Grill_2014,Pearson_2018,Typel_2010} \\
    GPPVA(DDME2) & 2.48 & 13.30 & 328.1 & No & \cite{Grill_2014,Pearson_2018,Lalazissis_2005} \\
    GPPVA(FSU2) & 2.07 & 13.71 & 346.8 & No & \cite{Providencia_2019,Grill_2014,Pearson_2018,Chen:2014sca} \\
    GPPVA(FSU2H) & 2.38 & 13.38 & 354.7 & No & \cite{Providencia_2019,Grill_2014,Pearson_2018,Negreiros2018} \\
    GPPVA(FSU2R) & 2.05 & 12.91 & 260.5 & No & \cite{Providencia_2019,Grill_2014,Pearson_2018,Negreiros2018} \\
    GPPVA(NL3wrL55) & 2.75 & 13.87 & 451.8 & No & \cite{Grill_2014,Pearson_2018,Pais:2016xiu} \\
    GPPVA(TM1e) & 2.12 & 13.14 & 292.3 & No & \cite{Grill_2014,Pearson_2018,Shen:2020sec} \\
    GPPVA(TW) & 2.08 & 12.18 & 161.0 & No & \cite{Grill_2014,Pearson_2018,Typel_1999} \\
    OPGR(DD2HdeltaY4) & 2.05 & 12.65 & 252.7 & Yes & \cite{Gaitanos_2004, oertel2014hyperonsneutronstarmatter, Grill_2014, Douchin_2001} \\
    OPGR(GM1Y5) & 2.12 & 13.77 & 425.6 & Yes & \cite{Glendenning_1991,Douchin_2001,oertel2014hyperonsneutronstarmatter} \\
    OPGR(GM1Y6) & 2.29 & 13.77 & 425.5 & Yes & \cite{Glendenning_1991,Douchin_2001,oertel2014hyperonsneutronstarmatter} \\
    PCGS(PCSB0) & 2.53 & 13.34 & 324.3 & No & \cite{Hempel_2010,Hornick_2021,Pradhan_2023} \\
    PCGS(PCSB1) & 2.19 & 13.04 & 269.8 & No & \cite{Hempel_2010,Pradhan_2023} \\
    PCGS(PCSB2) & 2.02 & 12.76 & 225.2 & No & \cite{Hempel_2010,Pradhan_2023} \\
    PCP(BSK22) & 2.26 & 12.98 & 266.3 & No & \cite{Goriely_2013,Xu_2013,Blumenfeld_2017,Wang_2017,Pearson_2018,Perot_2019,Pearson_2020,Allard_2021,Pearson_2022} \\
    PCP(BSK24) & 2.28 & 12.56 & 224.1 & No & \cite{Goriely_2013,Xu_2013,Blumenfeld_2017,Wang_2017,Pearson_2018,Perot_2019,Pearson_2020,Allard_2021,Pearson_2022} \\
    PCP(BSK25) & 2.22 & 12.39 & 210.2 & No & \cite{Goriely_2013,Xu_2013,Blumenfeld_2017,Wang_2017,Pearson_2018,Perot_2019,Pearson_2020,Allard_2021,Pearson_2022} \\
    PCP(BSK26) & 2.17 & 11.69 & 131.5 & No & \cite{Goriely_2013,Xu_2013,Blumenfeld_2017,Wang_2017,Pearson_2018,Perot_2019,Pearson_2020,Allard_2021,Pearson_2022} \\
    PT(GRDF2-DD2) & 2.42 & 13.20 & 309.8 & No & \cite{Typel_2010,Pais_2017,Typel_2018} \\
    R(DD2YDelta)\,1.1-1.1 & 2.04 & 12.84 & 232.9 & Yes & \cite{Typel_2010,Raduta_2020, Vi_as_2021, Raduta_2022} \\ %\,cold\,NS
    R(DD2YDelta)\,1.2-1.3 & 2.03 & 13.22 & 300.6 & Yes & \cite{Typel_2010,Raduta_2020, Vi_as_2021, Raduta_2022} \\
    R(DD2YDelta)\,1.2-1.1 & 2.05 & 12.22 & 166.3 & Yes & \cite{Typel_2010,Raduta_2020, Vi_as_2021, Raduta_2022} \\
    RG(KDE0v) & 1.96 & 11.19 & 86.9 & No & \cite{Gulminelli_2015,Agrawal_2005,Danielewicz_2009} \\
    RG(KDE0v1) & 1.97 & 11.38 & 96.1 & No & \cite{Gulminelli_2015,Agrawal_2005,Danielewicz_2009} \\
    RG(Rs) & 2.12 & 12.75 & 227.0 & No & \cite{Friedrich_1986,Gulminelli_2015,Danielewicz_2009} \\
    RG(SK255) & 2.14 & 12.93 & 227.0 & No & \cite{Gulminelli_2015,Agrawal_2003,Danielewicz_2009} \\
    RG(SK272) & 2.23 & 13.15 & 259.2 & No & \cite{Gulminelli_2015,Agrawal_2003,Danielewicz_2009} \\
    RG(SKI2) & 2.16 & 13.31 & 300.7 & No & \cite{Reinhard_1995,Gulminelli_2015,Danielewicz_2009} \\
    RG(SKa) & 2.21 & 12.79 & 228.0 & No & \cite{Deshmukh_1976,Gulminelli_2015,Danielewicz_2009} \\
    RG(SKb) & 2.19 & 12.18 & 192.5 & No & \cite{Deshmukh_1976,Gulminelli_2015,Danielewicz_2009} \\
    RG(SLY2) & 2.05 & 11.63 & 118.7 & No & \cite{Gulminelli_2015,Danielewicz_2009} \\
    RG(SLY2) & 2.10 & 11.73 & 131.2 & No & \cite{Gulminelli_2015,Danielewicz_2009} \\
    RG(SLY4) & 2.05 & 11.55 & 114.0 & No & \cite{Danielewicz_2009,Gulminelli_2015,Chabanat_1998} \\
    RG(SLY9) & 2.15 & 12.35 & 181.0 & No & \cite{Gulminelli_2015,Danielewicz_2009} \\
    RG(SkI3) & 2.24 & 13.44 & 321.0 & No & \cite{Reinhard_1995,Gulminelli_2015,Danielewicz_2009} \\
    RG(SkI4) & 2.17 & 12.31 & 191.0 & No & \cite{Reinhard_1995,Gulminelli_2015,Danielewicz_2009} \\
    RG(SkI5) & 2.24 & 13.91 & 400.9 & No & \cite{Reinhard_1995,Gulminelli_2015,Danielewicz_2009} \\
    RG(SkI6) & 2.19 & 12.42 & 201.4 & No & \cite{Reinhard_1995,Gulminelli_2015,Danielewicz_2009} \\
    RG(SkMp) & 2.11 & 12.34 & 184.8 & No & \cite{Rhoades_Brown_1989,Gulminelli_2015,Danielewicz_2009} \\
    RG(SkOp) & 1.97 & 11.86 & 129.4 & No & \cite{Lee_1988,Gulminelli_2015,Danielewicz_2009} \\
    SPG(M1) & 2.54 & 12.83 & 247.4 & No & \cite{Pearson_2018,Scurto:2024ekq} \\
    SPG(M2) & 2.42 & 12.65 & 234.8 & No &  \cite{Pearson_2018,Scurto:2024ekq}\\
    SPG(M3) & 2.68 & 12.75 & 249.8 & No &  \cite{Pearson_2018,Scurto:2024ekq}\\
    SPG(M4) & 2.35 & 12.31 & 194.3 & No &  \cite{Pearson_2018,Scurto:2024ekq}\\
    SPG(M5) & 2.70 & 13.50 & 365.4 & No &  \cite{Pearson_2018,Scurto:2024ekq}\\
    VGBCMR(D1M) & 2.01 & 11.58 & 122.4 & No & \cite{Gonzalez_Boquera_2018,Mondal_2020,Vi_as_2021} \\
    XMLSLZ(DD-LZ1) & 2.56 & 13.26 & 344.2 & No & \cite{Wei_2020,Xia_2022} \\
    XMLSLZ(DDMEX) & 2.56 & 13.46 & 369.3 & No & \cite{Taninah_2020,Xia_2022} \\
    XMLSLZ(DDME2) & 2.48 & 13.26 & 329.5 & No & \cite{Lalazissis_2005,Xia_2022} \\
    XMLSLZ(GM1) & 2.36 & 13.71 & 393.0 & No & \cite{Glendenning_1991,Xia_2022} \\
    XMLSLZ(MTVTC) & 2.02 & 12.83 & 230.4 & No & \cite{Maruyama_2005,Xia_2022} \\
    XMLSLZ(NL3) & 2.77 & 14.62 & 591.0 & No & \cite{Lalazissis_1997,Xia_2022} \\
    XMLSLZ(PK1) & 2.31 & 14.29 & 478.9 & No & \cite{Long_2004,Xia_2022} \\
    XMLSLZ(PKDD) & 2.33 & 13.54 & 326.8 & No & \cite{Long_2004,Xia_2022} \\
    XMLSLZ(TM1) & 2.18 & 14.14 & 437.8 & No & \cite{Sugahara_1994,Xia_2022} \\
    XMLSLZ(TW99) & 2.08 & 12.13 & 161.0 & No & \cite{Typel_1999,Xia_2022} \\
    eos1 & 2.21 & 13.08 & 341.9 & Yes & \cite{Char2025b,Davis:2024nda} \\
    eos2 & 1.94 & 14.15 & 564.8 & Yes & \cite{Char2025b,Davis:2024nda} \\
    eos3 & 2.06 & 14.12 & 479.2 & Yes & \cite{Char2025b,Davis:2024nda} \\
    eos4 & 2.07 & 13.45 & 363.9 & Yes & \cite{Char2025b,Davis:2024nda} \\
    eos5 & 2.07 & 13.53 & 399.8 & Yes & \cite{Char2025b,Davis:2024nda} \\
    eos6 & 1.92 & 13.43 & 334.7 & Yes & \cite{Char2025b,Davis:2024nda} \\
    eos7 & 2.03 & 13.10 & 295.9 & Yes & \cite{Char2025b,Davis:2024nda} \\
    eos9 & 2.00 & 12.65 & 222.2 & Yes & \cite{Char2025b,Davis:2024nda} \\
    eos0 & 2.18 & 13.19 & 342.4 & Yes & \cite{Char2025b,Davis:2024nda} \\
    eos1 & 2.08 & 13.07 & 339.5 & Yes & \cite{Char2025b,Davis:2024nda} \\
    eos2 & 2.35 & 14.15 & 564.8 & Yes & \cite{Char2025b,Davis:2024nda} \\
    eos3 & 2.07 & 14.12 & 478.3 & Yes & \cite{Char2025b,Davis:2024nda} \\
    eos4 & 2.08 & 13.45 & 363.4 & Yes & \cite{Char2025b,Davis:2024nda} \\
    eos5 & 2.08 & 13.52 & 398.1 & Yes & \cite{Char2025b,Davis:2024nda} \\
    eos6 & 1.97 & 13.45 & 340.4 & Yes & \cite{Char2025b,Davis:2024nda} \\
    eos8 & 2.16 & 13.06 & 304.1 & Yes & \cite{Char2025b,Davis:2024nda} \\
    eos0 & 2.56 & 13.19 & 342.4 & No & \cite{Char_2023,Davis:2024nda} \\
    eos7 & 2.29 & 13.10 & 295.8 & No & \cite{Char_2023,Davis:2024nda} \\
    (GDFM)eos1 & 2.36 & 12.89 & 292.7 & No & \cite{Char_2023,Davis:2024nda} \\
    (GDFM)eos3 & 2.30 & 12.63 & 271.9 & No & \cite{Char_2023,Davis:2024nda} \\
    (GDFM)eos6 & 2.26 & 12.75 & 240.3 & No & \cite{Char_2023,Davis:2024nda} \\
    (GDFM)eos8 & 2.51 & 14.38 & 549.8 & No & \cite{Char_2023,Davis:2024nda} \\
    (GDFM)eos11 & 2.92 & 14.18 & 510.1 & No & \cite{Char_2023,Davis:2024nda} \\
    (GDFM)eos12 & 2.45 & 13.70 & 372.3 & No & \cite{Char_2023,Davis:2024nda} \\
    (GDFM)eos15 & 2.72 & 14.17 & 446.5 & No & \cite{Char_2023,Davis:2024nda} \\
    (GDFM)eos19 & 2.27 & 12.74 & 238.4 & No & \cite{Char_2023,Davis:2024nda} \\
    (GDFM)eos20 & 2.65 & 13.63 & 384.4 & No & \cite{Char_2023,Davis:2024nda} \\
    (GDFM)eos22 & 2.74 & 14.94 & 629.6 & No & \cite{Char_2023,Davis:2024nda} \\
    (GDFM)eos24 & 2.51 & 14.15 & 401.3 & No & \cite{Char_2023,Davis:2024nda} \\
    (GDFM)eos28 & 2.18 & 12.69 & 214.1 & No & \cite{Char_2023,Davis:2024nda} \\
    (GDFM)eos30 & 2.87 & 15.16 & 624.8 & No & \cite{Char_2023,Davis:2024nda} \\
    (GDFM)eos36 & 2.46 & 13.57 & 351.3 & No & \cite{Char_2023,Davis:2024nda} \\
    (GDFM)eos39 & 2.51 & 14.65 & 583.8 & No & \cite{Char_2023,Davis:2024nda} \\
    (GDFM)eos41 & 2.52 & 13.76 & 428.5 & No & \cite{Char_2023,Davis:2024nda} \\
    (GDFM)eos43 & 2.77 & 13.86 & 449.0 & No & \cite{Char_2023,Davis:2024nda} \\
    (GDFM)eos45 & 2.77 & 13.74 & 414.8 & No & \cite{Char_2023,Davis:2024nda} \\
    (GDFM)eos48 & 2.19 & 13.11 & 264.9 & No & \cite{Char_2023,Davis:2024nda} \\
    (GDFM)eos50 & 2.18 & 12.87 & 201.1 & No & \cite{Char_2023,Davis:2024nda} \\
    (GDFM)eos52 & 2.08 & 11.66 & 135.8 & No & \cite{Char_2023,Davis:2024nda} \\
    (GDFM)eos58 & 2.65 & 13.87 & 424.6 & No & \cite{Char_2023,Davis:2024nda} \\
    (GDFM)eos60 & 2.63 & 14.06 & 456.8 & No & \cite{Char_2023,Davis:2024nda} \\
    (GDFM)eos61 & 2.47 & 14.30 & 448.7 & No & \cite{Char_2023,Davis:2024nda} \\
    (GDFM)eos64 & 2.18 & 12.99 & 282.7 & No & \cite{Char_2023,Davis:2024nda} \\
    (GDFM)eos66 & 2.74 & 14.87 & 528.8 & No & \cite{Char_2023,Davis:2024nda} \\
    (GDFM)eos67 & 2.73 & 14.28 & 552.6 & No & \cite{Char_2023,Davis:2024nda} \\
    (GDFM)eos72 & 2.88 & 15.18 & 642.1 & No & \cite{Char_2023,Davis:2024nda} \\
    (GDFM)eos78 & 2.88 & 14.53 & 621.8 & No & \cite{Char_2023,Davis:2024nda} \\
    (GDFM)eos80 & 2.49 & 13.82 & 391.5 & No & \cite{Char_2023,Davis:2024nda} \\
    (GDFM)eos82 & 2.77 & 14.12 & 540.5 & No & \cite{Char_2023,Davis:2024nda} \\
    (GDFM)eos84 & 2.83 & 15.03 & 625.8 & No & \cite{Char_2023,Davis:2024nda} \\
    (GDFM)eos85 & 2.63 & 15.02 & 593.5 & No & \cite{Char_2023,Davis:2024nda} \\
    (GDFM)eos89 & 2.19 & 13.44 & 301.4 & No & \cite{Char_2023,Davis:2024nda} \\
    (GDFM)eos92 & 2.77 & 14.61 & 584.4 & No & \cite{Char_2023,Davis:2024nda} \\
    (GDFM)eos94 & 2.24 & 12.51 & 241.1 & No & \cite{Char_2023,Davis:2024nda} \\
    (GDFM)eos10 & 2.55 & 14.09 & 479.2 & No & \cite{Char_2023,Davis:2024nda} \\
    (GDFM)eos17 & 2.74 & 13.41 & 346.0 & No & \cite{Char_2023,Davis:2024nda} \\
    (GDFM)eos54 & 2.44 & 12.88 & 304.4 & No & \cite{Char_2023,Davis:2024nda} \\
    (GDFM)eos73 & 2.37 & 14.05 & 455.6 & No & \cite{Char_2023,Davis:2024nda} \\
    (GDFM)eos99 & 2.33 & 12.32 & 233.0 & No & \cite{Char_2023,Davis:2024nda} \\
    eos101 & 2.59 & 13.72 & 422.5 & No & \cite{Char_2023,Davis:2024nda} \\
    eos102 & 2.47 & 12.77 & 301.0 & No & \cite{Char_2023,Davis:2024nda} \\
    eos103 & 2.53 & 13.68 & 420.6 & No & \cite{Char_2023,Davis:2024nda} \\
    eos105 & 2.59 & 13.16 & 327.7 & No & \cite{Char_2023,Davis:2024nda} \\
    eos106 & 2.62 & 13.40 & 381.4 & No & \cite{Char_2023,Davis:2024nda} \\
    eos108 & 2.59 & 13.29 & 366.8 & No & \cite{Char_2023,Davis:2024nda} \\
    eos109 & 2.75 & 13.85 & 454.9 & No & \cite{Char_2023,Davis:2024nda} \\
    eos110 & 2.39 & 13.02 & 277.4 & No & \cite{Char_2023,Davis:2024nda} \\
    eos111 & 2.32 & 13.31 & 355.5 & No & \cite{Char_2023,Davis:2024nda} \\
    eos112 & 2.64 & 13.16 & 311.6 & No & \cite{Char_2023,Davis:2024nda} \\
    eos114 & 2.38 & 13.21 & 311.5 & No & \cite{Char_2023,Davis:2024nda} \\
    eos117 & 2.35 & 12.65 & 254.3 & No & \cite{Char_2023,Davis:2024nda} \\
    eos118 & 2.50 & 13.58 & 382.3 & No & \cite{Char_2023,Davis:2024nda} \\
    eos119 & 2.40 & 13.30 & 335.8 & No & \cite{Char_2023,Davis:2024nda} \\
    eos122 & 2.61 & 14.28 & 566.2 & No & \cite{Char_2023,Davis:2024nda} \\
    eos125 & 2.41 & 12.62 & 289.9 & No & \cite{Char_2023,Davis:2024nda} \\
    eos126 & 2.74 & 14.32 & 551.9 & No & \cite{Char_2023,Davis:2024nda} \\
    eos128 & 2.54 & 13.58 & 423.9 & No & \cite{Char_2023,Davis:2024nda} \\
    eos13 & 2.57 & 14.05 & 532.0 & No & \cite{Char_2023,Davis:2024nda} \\
    eos130 & 2.69 & 13.65 & 400.9 & No & \cite{Char_2023,Davis:2024nda} \\
    eos131 & 2.60 & 14.27 & 571.5 & No & \cite{Char_2023,Davis:2024nda} \\
    eos132 & 2.42 & 12.66 & 248.3 & No & \cite{Char_2023,Davis:2024nda} \\
    eos133 & 2.40 & 13.33 & 349.4 & No & \cite{Char_2023,Davis:2024nda} \\
    eos135 & 2.68 & 13.91 & 441.6 & No & \cite{Char_2023,Davis:2024nda} \\
    eos138 & 2.88 & 13.80 & 506.3 & No & \cite{Char_2023,Davis:2024nda} \\
    eos14 & 2.64 & 13.89 & 483.4 & No & \cite{Char_2023,Davis:2024nda} \\
    eos144 & 2.32 & 12.59 & 250.0 & No & \cite{Char_2023,Davis:2024nda} \\
    eos147 & 2.51 & 12.76 & 266.5 & No & \cite{Char_2023,Davis:2024nda} \\
    eos149 & 2.42 & 12.70 & 275.2 & No & \cite{Char_2023,Davis:2024nda} \\
    eos158 & 2.53 & 14.06 & 489.1 & No & \cite{Char_2023,Davis:2024nda} \\
    eos16 & 2.60 & 13.85 & 471.4 & No & \cite{Char_2023,Davis:2024nda} \\
    eos160 & 2.72 & 13.27 & 374.9 & No & \cite{Char_2023,Davis:2024nda} \\
    eos161 & 2.22 & 12.74 & 243.9 & No & \cite{Char_2023,Davis:2024nda} \\
    eos162 & 2.43 & 12.94 & 317.0 & No & \cite{Char_2023,Davis:2024nda} \\
    eos166 & 2.47 & 13.57 & 403.9 & No & \cite{Char_2023,Davis:2024nda} \\
    eos168 & 2.53 & 14.13 & 499.3 & No & \cite{Char_2023,Davis:2024nda} \\
    eos171 & 2.41 & 12.61 & 266.9 & No & \cite{Char_2023,Davis:2024nda} \\
    eos173 & 2.64 & 13.11 & 379.5 & No & \cite{Char_2023,Davis:2024nda} \\
    eos175 & 2.21 & 12.25 & 198.8 & No & \cite{Char_2023,Davis:2024nda} \\
    eos176 & 2.47 & 12.86 & 311.1 & No & \cite{Char_2023,Davis:2024nda} \\
    eos178 & 2.30 & 12.69 & 251.1 & No & \cite{Char_2023,Davis:2024nda} \\
    eos179 & 2.59 & 13.28 & 370.2 & No & \cite{Char_2023,Davis:2024nda} \\
    eos18 & 2.77 & 13.33 & 421.0 & No & \cite{Char_2023,Davis:2024nda} \\
    eos182 & 2.79 & 13.41 & 432.4 & No & \cite{Char_2023,Davis:2024nda} \\
    eos186 & 2.68 & 13.19 & 367.0 & No & \cite{Char_2023,Davis:2024nda} \\
    eos188 & 2.47 & 13.26 & 316.2 & No & \cite{Char_2023,Davis:2024nda} \\
    eos191 & 2.46 & 14.27 & 536.7 & No & \cite{Char_2023,Davis:2024nda} \\
    eos193 & 2.52 & 13.46 & 363.5 & No & \cite{Char_2023,Davis:2024nda} \\
    eos196 & 2.48 & 12.75 & 303.7 & No & \cite{Char_2023,Davis:2024nda} \\
    eos198 & 2.48 & 12.79 & 297.0 & No & \cite{Char_2023,Davis:2024nda} \\
    eos2 & 2.67 & 14.15 & 564.8 & No & \cite{Char_2023,Davis:2024nda} \\
    eos200 & 2.55 & 13.68 & 395.5 & No & \cite{Char_2023,Davis:2024nda} \\
    eos203 & 2.31 & 13.39 & 380.6 & No & \cite{Char_2023,Davis:2024nda} \\
    eos207 & 2.52 & 14.08 & 504.9 & No & \cite{Char_2023,Davis:2024nda} \\
    eos208 & 2.59 & 13.16 & 354.5 & No & \cite{Char_2023,Davis:2024nda} \\
    eos211 & 2.43 & 13.05 & 331.3 & No & \cite{Char_2023,Davis:2024nda} \\
    eos213 & 2.59 & 14.12 & 510.3 & No & \cite{Char_2023,Davis:2024nda} \\
    eos216 & 2.75 & 13.46 & 404.7 & No & \cite{Char_2023,Davis:2024nda} \\
    eos223 & 2.42 & 13.51 & 351.3 & No & \cite{Char_2023,Davis:2024nda} \\
    eos224 & 2.63 & 14.06 & 496.5 & No & \cite{Char_2023,Davis:2024nda} \\
    eos23 & 2.46 & 13.58 & 371.6 & No & \cite{Char_2023,Davis:2024nda} \\
    eos231 & 2.46 & 13.50 & 373.9 & No & \cite{Char_2023,Davis:2024nda} \\
    eos232 & 2.65 & 13.05 & 322.3 & No & \cite{Char_2023,Davis:2024nda} \\
    eos234 & 2.52 & 12.72 & 296.1 & No & \cite{Char_2023,Davis:2024nda} \\
    eos26 & 2.30 & 13.38 & 341.7 & No & \cite{Char_2023,Davis:2024nda} \\
    eos27 & 2.67 & 13.36 & 416.8 & No & \cite{Char_2023,Davis:2024nda} \\
    eos29 & 2.78 & 13.49 & 397.3 & No & \cite{Char_2023,Davis:2024nda} \\
    eos33 & 2.18 & 13.15 & 294.6 & No & \cite{Char_2023,Davis:2024nda} \\
    eos35 & 2.35 & 13.68 & 387.3 & No & \cite{Char_2023,Davis:2024nda} \\
    eos37 & 2.40 & 13.73 & 411.3 & No & \cite{Char_2023,Davis:2024nda} \\
    eos4 & 2.42 & 13.46 & 365.1 & No & \cite{Char_2023,Davis:2024nda} \\
    eos42 & 2.41 & 12.61 & 300.6 & No & \cite{Char_2023,Davis:2024nda} \\
    eos46 & 2.29 & 13.25 & 348.6 & No & \cite{Char_2023,Davis:2024nda} \\
    eos47 & 2.37 & 13.08 & 308.5 & No & \cite{Char_2023,Davis:2024nda} \\
    eos49 & 2.48 & 14.08 & 513.5 & No & \cite{Char_2023,Davis:2024nda} \\
    eos5 & 2.47 & 13.53 & 400.4 & No & \cite{Char_2023,Davis:2024nda} \\
    eos53 & 2.43 & 14.17 & 492.0 & No & \cite{Char_2023,Davis:2024nda} \\
    eos54 & 2.37 & 12.73 & 288.5 & No & \cite{Char_2023,Davis:2024nda} \\
    eos55 & 2.76 & 13.83 & 431.2 & No & \cite{Char_2023,Davis:2024nda} \\
    eos56 & 2.46 & 12.53 & 268.3 & No & \cite{Char_2023,Davis:2024nda} \\
    eos57 & 2.50 & 13.36 & 348.4 & No & \cite{Char_2023,Davis:2024nda} \\
    eos59 & 2.69 & 13.31 & 362.4 & No & \cite{Char_2023,Davis:2024nda} \\
    eos63 & 2.38 & 13.65 & 376.7 & No & \cite{Char_2023,Davis:2024nda} \\
    eos65 & 2.73 & 13.20 & 377.1 & No & \cite{Char_2023,Davis:2024nda} \\
    eos68 & 2.24 & 12.55 & 227.3 & No & \cite{Char_2023,Davis:2024nda} \\
    eos69 & 2.76 & 14.18 & 566.6 & No & \cite{Char_2023,Davis:2024nda} \\
    eos71 & 2.43 & 12.93 & 310.7 & No & \cite{Char_2023,Davis:2024nda} \\
    eos76 & 2.32 & 12.67 & 258.0 & No & \cite{Char_2023,Davis:2024nda} \\
    eos77 & 2.37 & 12.70 & 273.8 & No & \cite{Char_2023,Davis:2024nda} \\
    eos78 & 2.85 & 13.42 & 435.4 & No & \cite{Char_2023,Davis:2024nda} \\
    eos79 & 2.29 & 12.87 & 248.7 & No & \cite{Char_2023,Davis:2024nda} \\
    eos81 & 2.86 & 13.88 & 472.0 & No & \cite{Char_2023,Davis:2024nda} \\
    eos83 & 2.99 & 14.22 & 595.6 & No & \cite{Char_2023,Davis:2024nda} \\
    eos87 & 2.73 & 14.36 & 577.6 & No & \cite{Char_2023,Davis:2024nda} \\
    eos88 & 2.68 & 13.27 & 385.3 & No & \cite{Char_2023,Davis:2024nda} \\
    eos9 & 2.22 & 12.68 & 227.5 & No & \cite{Char_2023,Davis:2024nda} \\
    eos91 & 2.42 & 13.38 & 383.4 & No & \cite{Char_2023,Davis:2024nda} \\
    eos93 & 2.54 & 12.97 & 345.4 & No & \cite{Char_2023,Davis:2024nda} \\
    eos97 & 2.86 & 14.01 & 512.0 & No & \cite{Char_2023,Davis:2024nda} \\
    eos98 & 2.73 & 14.11 & 545.6 & No & \cite{Char_2023,Davis:2024nda}\\
    \hline\hline
\end{longtable}
\end{center}

\begin{figure*}%[tb]
	\includegraphics[width=6cm]{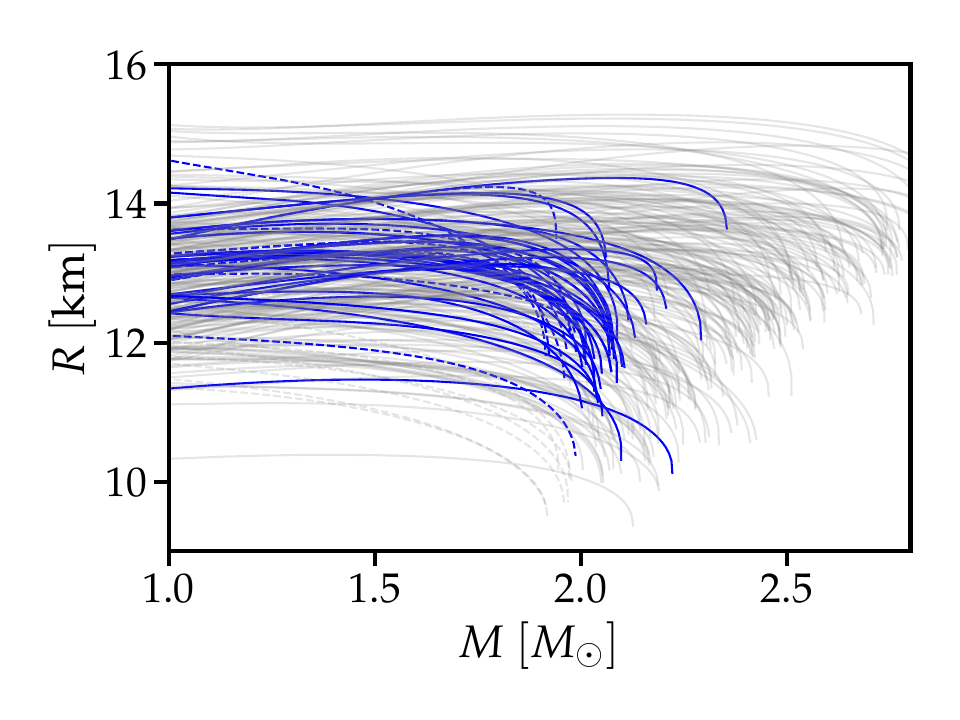}%Tovanalysis-discrete-RM.py
    \includegraphics[width=6cm]{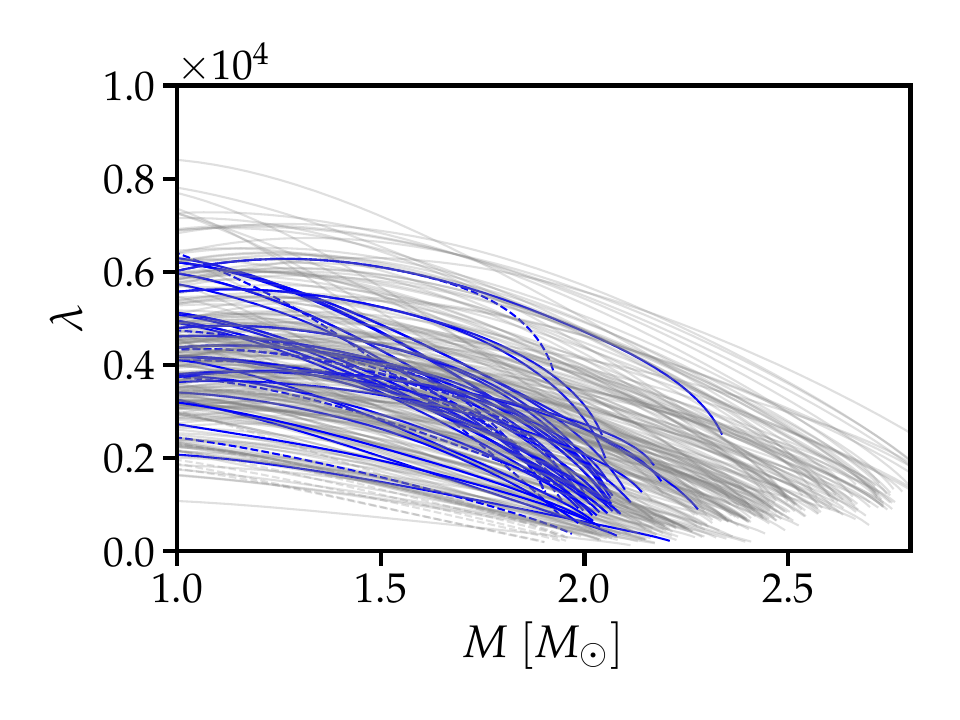}%Tovanalysis-discrete-Lambda-M.py
    \includegraphics[width=6cm]{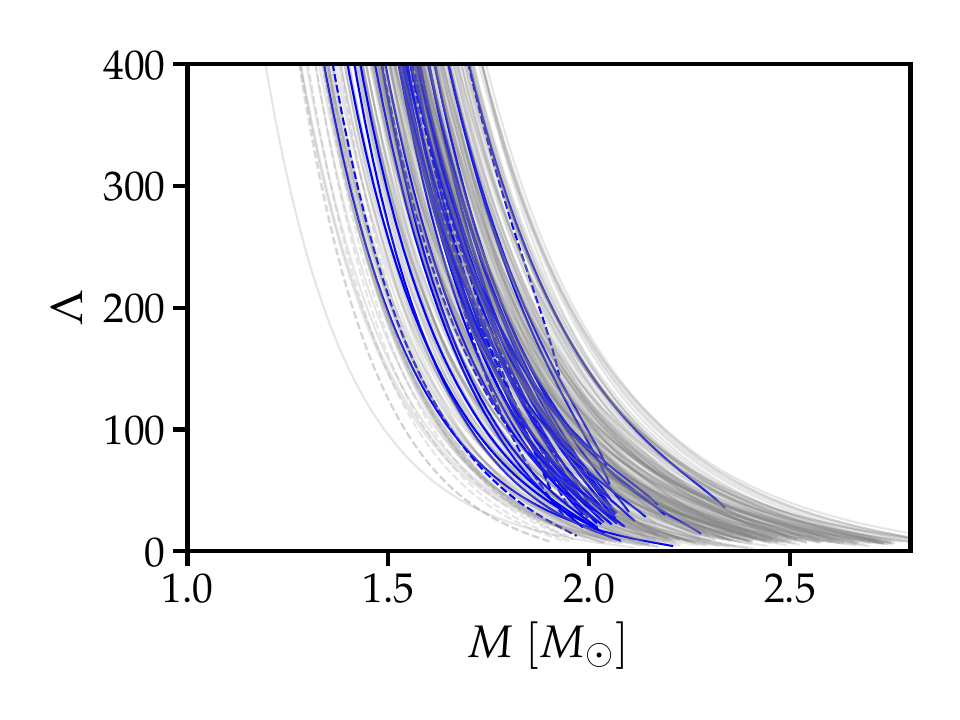}\\%Tovanalysis-discrete-capLambda-M.py
	\includegraphics[width=6cm]{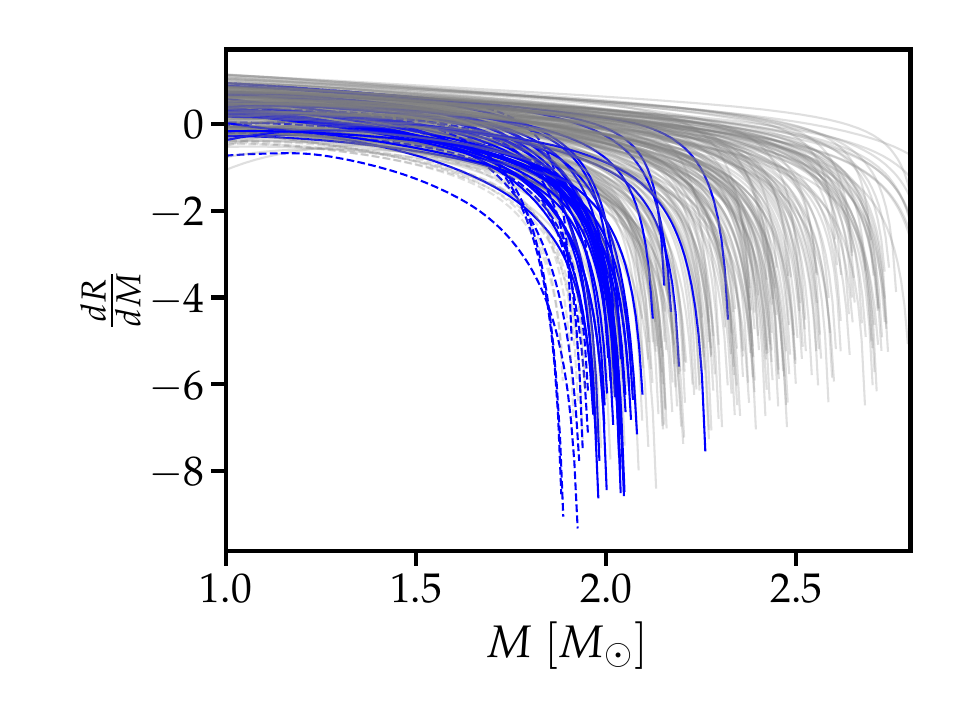}%Tovanalysis-discrete-slopeR-M.py
    \includegraphics[width=6cm]{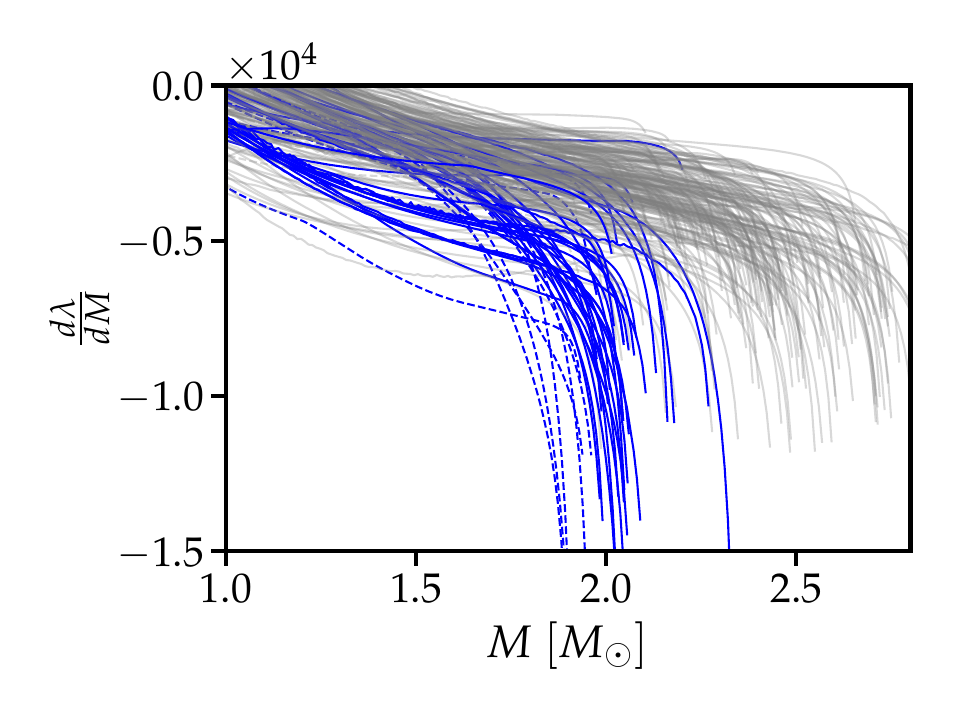}%Tovanalysis-discrete-slopeLambda-M.py
    \includegraphics[width=6cm]{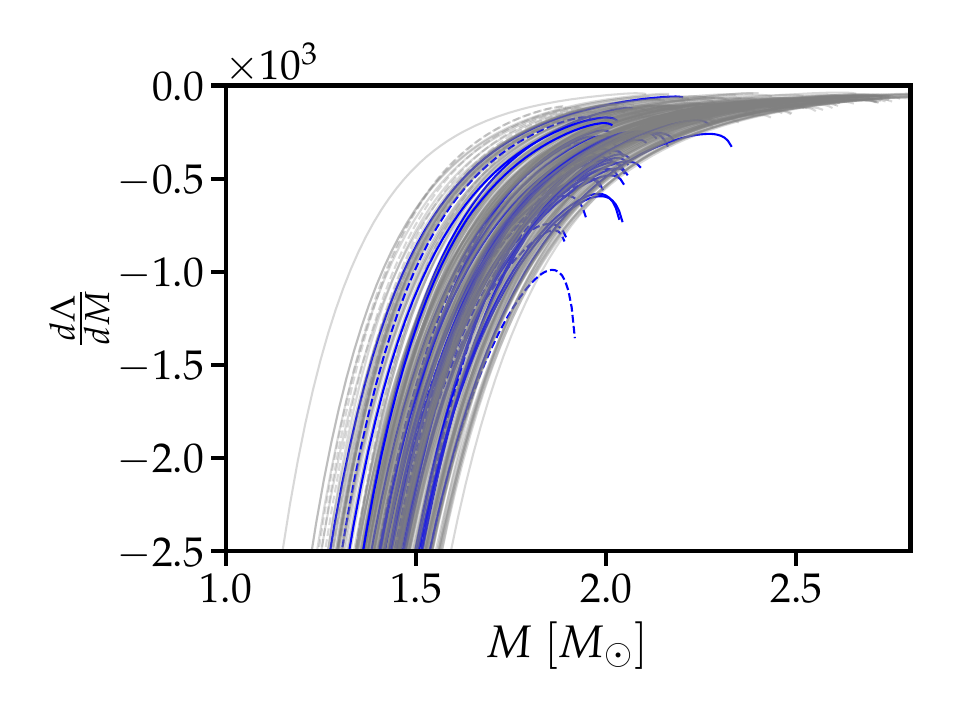}\\%Tovanalysis-discrete-slopecapLambda-M.py
	\includegraphics[width=6cm]{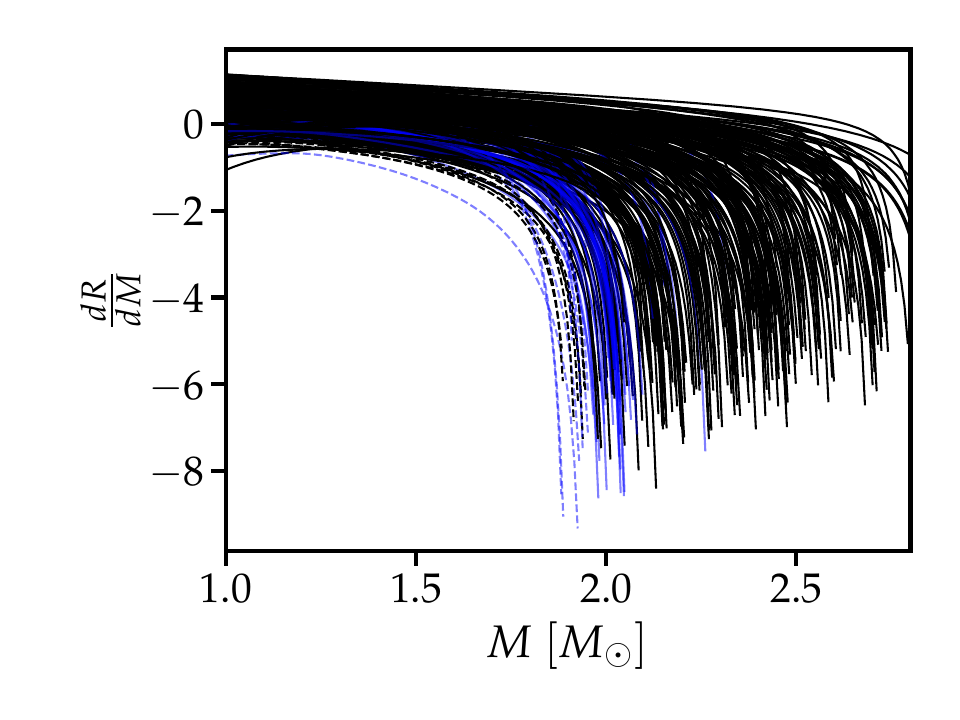}%Tovanalysis-discrete-slopeR-M-nucleonic-pronounced.py
    \includegraphics[width=6cm]{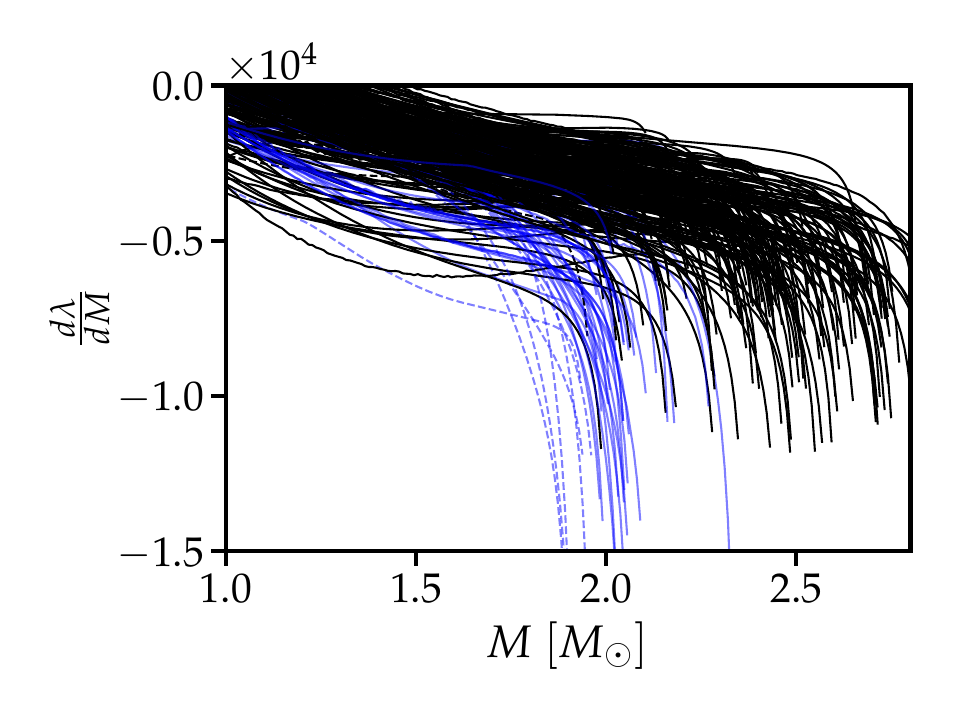}%Tovanalysis-discrete-slopeLambda-M-nucleonic-pronounced.py
    \includegraphics[width=6cm]{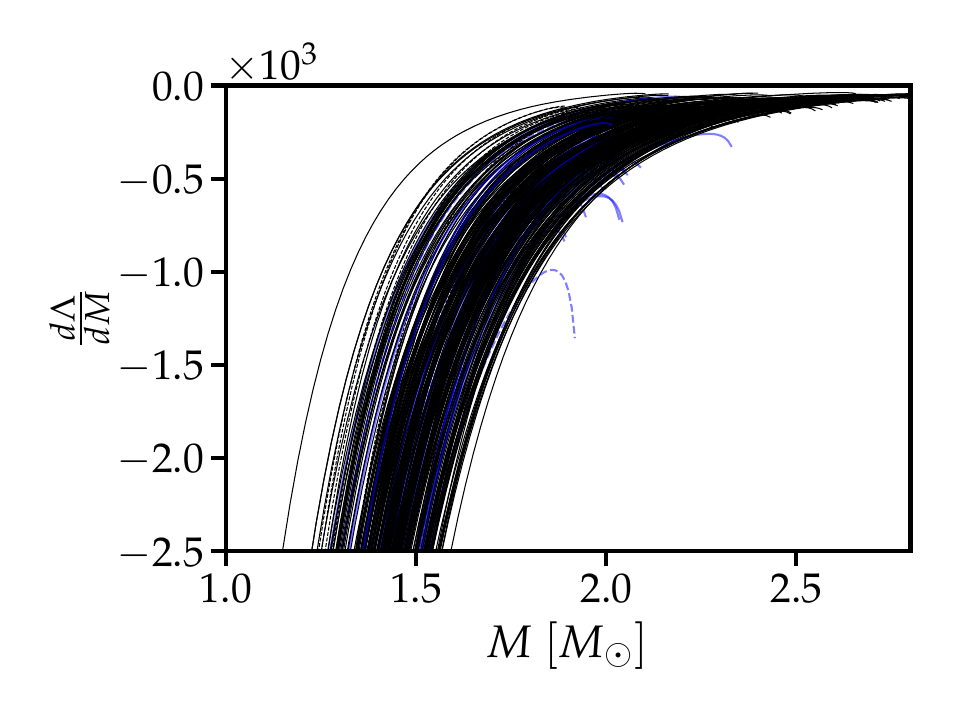}\\%Tovanalysis-discrete-slopecapLambda-M-nucleonic-pronounced.py
    \includegraphics[width=6cm]{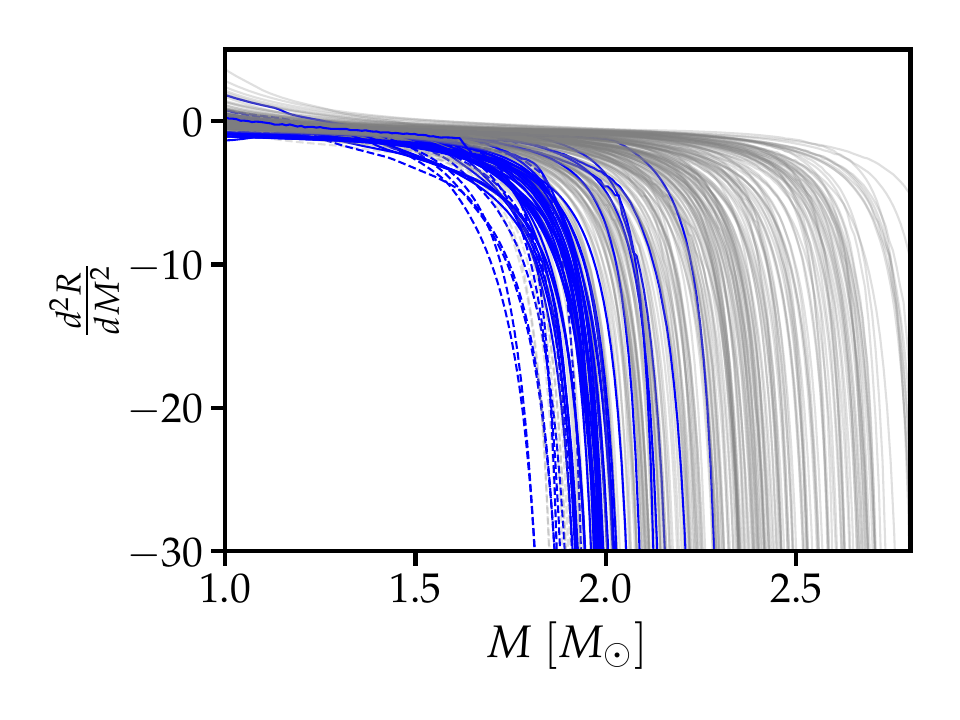}%Tovanalysis-discrete-secderivR-M.py
    \includegraphics[width=6cm]{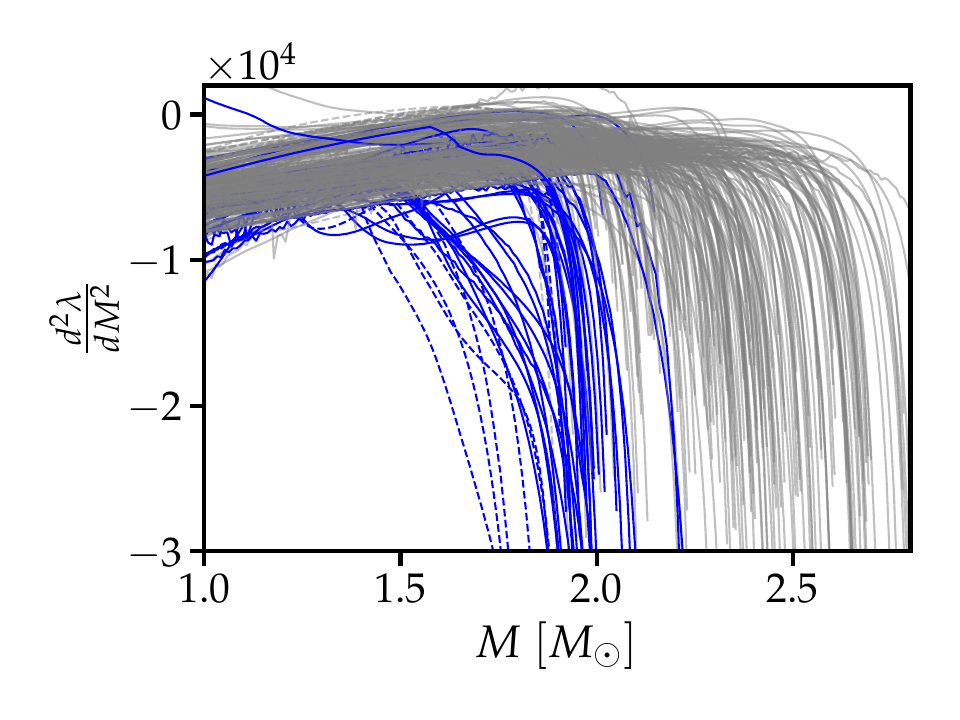}%Tovanalysis-discrete-secderivLambda-M-2.py
    \includegraphics[width=6cm]{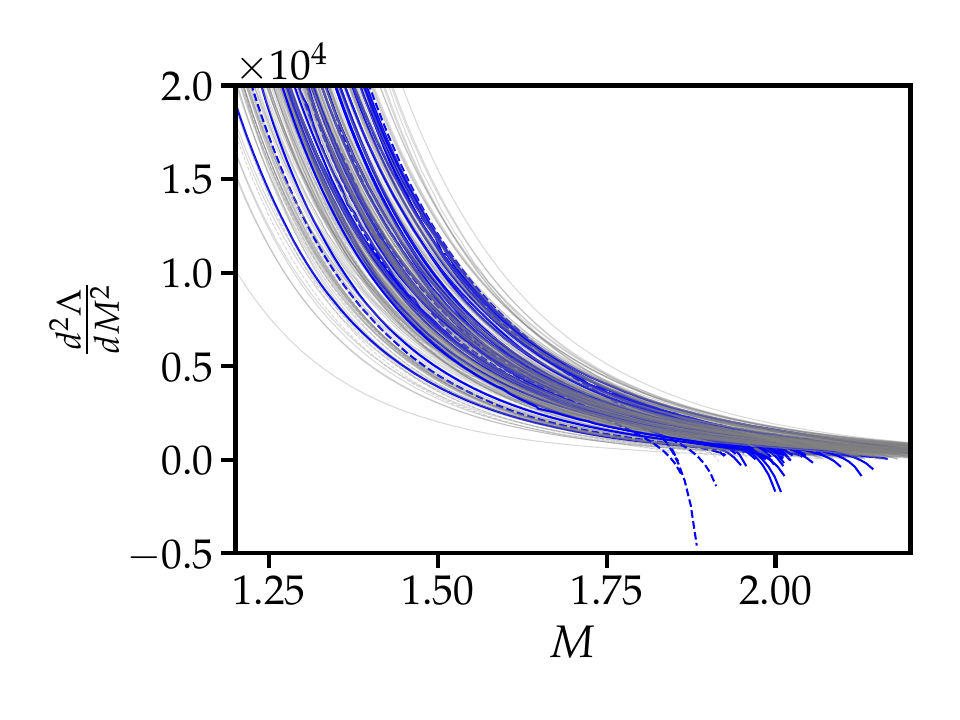} %   Tovanalysis-discrete-secderivcapLambda-M.py
    
    \caption{Upper panels: Stellar parameters of NSs as function of mass with the radius (left), the tidal deformabilty $\lambda$ (middle) and the mass-scaled tidal deformabilty $\Lambda=\lambda/M^5$ (right). The latter two quantities are dimensionless or in geometric units, respectively. Purely nucleonic models displayed in gray, while models with hyperons are highlighted in blue. Dashed curves indicate models with $M_\mathrm{max}<2.0~M_\odot$. Second-row panels: First derivatives with respect to mass of the respective quantity of the panel above. The same line scheme as in the upper panels is applied. Third-row panels: Same as second row but with purely nucleonic models visually more emphasized by black curves to assess overlap between hyperonic and purely nucleonic EoSs. Lower panels: Second derivatives with respect to mass of the respective quantity of the upper panel.}
    \label{fig:overview}
\end{figure*}

The solutions from the stellar structure equations~\cite{Tolman1939,Oppenheimer1939,Hinderer2008,Binnington2009,Hinderer2010} determining $R$, $\lambda$ and $\Lambda=\lambda/M^5$ are shown in Fig.~\ref{fig:overview}. If necessary we extend the EoS tables to smaller densities by adding a crust EoS~\cite{Douchin2001,Baym1971} such that the integration terminates at $P\approx3\times10^{14}\mathrm{dyne/cm^2}$ for all models\footnote{The matching between crust and core EoS can affect radii and tidal deformability and in principle the use of unified EoS models is desirable~\cite{Fortin2016,Perot2020,Suleiman2021,Fantina2022}. For considering derivatives especially at higher masses such effects are however less relevant.}. The figure also provides the first and second derivatives with respect to mass. We obtain these derivatives by numerical differentiation using a centered second-order finite difference formula with variable step size. We note that numerical values of the derivatives or $\kappa_R$ (see main text) can be noisy depending on the details and accuracy of the TOV solutions, which can be cured by using larger step sized for the finite differences (without effects on the smooth behavior of the curves). The panels in the third row display data which are identical to those in the second row but using a different color scheme to better visualize different purely nucleonic models. By this the overlap between hyperonic and purely nucleonic models becomes more apparent.

\begin{figure*}

    \includegraphics[width=8.4cm]{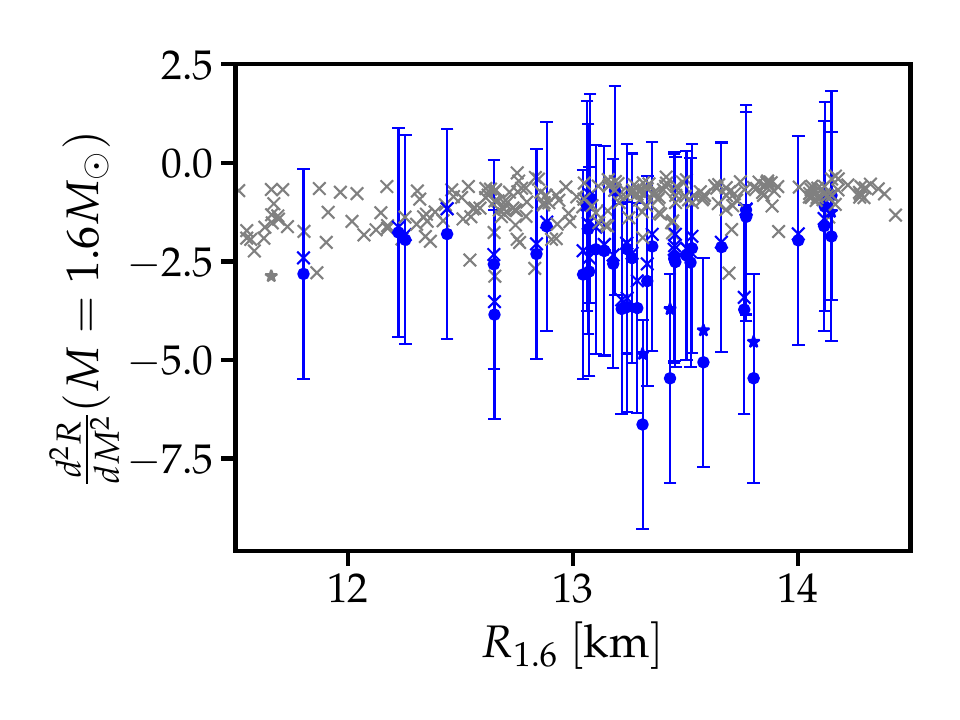} %   Tovanalysis-discrete-secderivR-M-mref-errors.py
    \includegraphics[width=8.4cm]{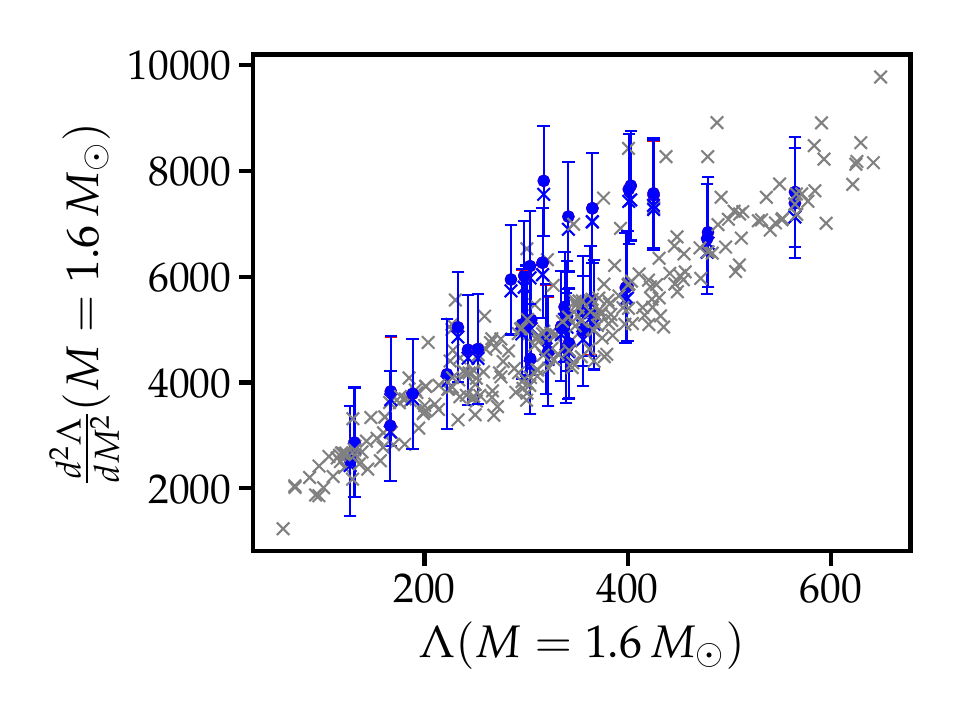} %   Tovanalysis-discrete-secderivcapLambda-M-mref-errors.py
    \caption{Second derivative of $R(M)$ (left) and $\Lambda(M)$ (right) with respect to mass $M$ at a fixed reference mass $M=1.6\,M_\odot$ as function of $R(M=1.6\,M_\odot)$ (left) and $\Lambda(M=1.6\,M_\odot)$, respectively. Symbols have the same meaning as in Fig.~3 of the main paper. We adopt $\delta R=100$~m and $\Delta M=0.25\,M_\odot$ for the left panel and $\delta \Lambda=100/1.6^5$ and $\Delta M=0.15\,M_\odot$ for the right panel. See main text for more details. }
    \label{fig:dr2dm2ref}
\end{figure*}

Figure~\ref{fig:dr2dm2ref} shows $\frac{d^2R}{dM^2}(M)$ and $\frac{d^2\Lambda}{dM^2}(M)$ at a fixed reference mass $M=1.6\,M_\odot$ similar to Figs.~3 and~5 in the main paper. For the right panel we adopt $\Delta=0.15~M_\odot$ because the discretization errors become comparable to errors stemming from the finite precision in $\Lambda$ if $\Delta=0.25~M_\odot$ is used.

%Control: production of eprint (1) enabled
%

%\bibliography{references.bib}

\end{document}